\documentclass[11pt]{article}
\usepackage[a4paper,margin=1in]{geometry}
\usepackage[T1]{fontenc}

\usepackage{amsopn,amsfonts,amsmath,amssymb,bm,dsfont}
\usepackage{graphicx,xurl,array,multirow}
\usepackage{algorithm,algorithmic}
\usepackage{subfig,caption,placeins,microtype}
\usepackage[numbers,sort&compress]{natbib}
\usepackage[hidelinks]{hyperref}
\newcommand{\paperabstract}{}
\newcommand{\abstractcontent}[1]{\gdef\paperabstract{#1}}
\newcommand{\paperkeywords}{}
\newcommand{\keyword}[1]{\gdef\paperkeywords{#1}}

\title{\bfseries
\makebox[\textwidth][c]{Entropy-Based and Quadratic Conditional Innovation Metrics}\\
\makebox[\textwidth][c]{for Informed Dynamic Scheduling of Belief-Propagation}\\
\makebox[\textwidth][c]{Decoding for LDPC Codes}}
\author{%
\normalsize Tofar C.-Y. Chang$^{1}$, Chia-En Liu$^{2}$, Ying-Ren Jian$^{2}$, and Di-Chun Liang$^{2}$\\[0.75em]
\small $^{1}$ Department of Electrical Engineering, National Taipei University of Technology,\\
\small Taipei, Taiwan\\
\small $^{2}$ Institute of Communications Engineering, National Yang Ming Chiao Tung University,\\
\small Hsinchu, Taiwan\\[0.4em]
}
\date{}
\abstractcontent{
We propose two enhanced conditional innovation (CI) metrics
for informed dynamic scheduling (IDS) in belief-propagation (BP) decoding
of LDPC codes.
The CI-based residual BP (CIRBP) algorithm uses the CI metric
to guide the selection of the variable node (VN) to be updated
and has been shown to outperform several existing IDS-based BP decoders,
especially at the early decoding stage.
However, the original CI metric uses only the absolute difference
between the current and precomputed posterior probabilities
and does not separately account for the reliability
associated with each probability.
By reformulating the CI metric relative to the decision boundary,
we separate each posterior probability into its decision side
and a linear-distance term reflecting reliability.
Replacing the latter with the binary entropy complement yields
the proposed entropy-based CI (E-CI) metric.
Moreover, the proposed quadratic CI (Q-CI) metric is derived
from the leading quadratic term in the local expansion
of the entropy complement,
avoiding the logarithmic operations required for direct E-CI evaluation.
We apply both proposed metrics to the CIRBP decoder
and its node-wise variant.
Numerical results show that the proposed metrics
generally provide lower frame error rates than the original CI metric,
with their relative advantages depending on
the code and decoding procedure.
}

\keyword{
belief propagation; binary entropy; conditional innovation; informed dynamic scheduling; LDPC codes
} 
\begin{document}
\maketitle
\begin{abstract}
\paperabstract
\end{abstract}

\section{Introduction}
\label{Sec1}

Low-density parity-check (LDPC) codes can achieve
capacity-approaching error-rate performance
under belief-propagation (BP) decoding \cite{LDPC}
and have been adopted in many communication standards,
such as IEEE 802.11 \cite{WiFi},
fifth-generation (5G) New Radio (NR) \cite{NR},
and DVB-S2 \cite{DVBS2}.
The conventional BP decoder, also known as the sum-product algorithm,
typically employs a flooding schedule on the associated Tanner graph,
in which variable nodes (VNs) and check nodes (CNs)
iteratively exchange variable-to-check (V2C) and check-to-variable (C2V) messages
in alternating parallel phases \cite{LDPC}.
Although the flooding schedule offers a high degree of parallelism,
it usually requires several decoding iterations to converge
and may entail complex interconnections and substantial memory requirements
for hardware implementation.
To address these limitations,
sequential and partially sequential schedules have been proposed.
By processing only part of the code graph at a time,
these schedules allow newly updated messages
to be used in subsequent computations
within the same decoding iteration.
Two basic forms are layered BP (LBP) decoding,
which sequentially updates individual CNs or groups of CNs \cite{LBP},
and shuffled BP (SBP) decoding,
which similarly updates individual VNs or groups of VNs \cite{SBP}.
Various node grouping methods
based on node and graph characteristics
and message processing orders
have subsequently been developed for these two scheduling schemes,
with the processing order either predetermined
or adapted according to information generated during iterative decoding
\cite{newGHSBP,Row_Layer,LPHD_RB-LBP,LG-GSBP,AGSBP,WRB-LBP}.

Informed dynamic scheduling (IDS)
is a family of adaptive scheduling strategies
for BP decoding that has been widely studied in the literature.
An IDS strategy uses information available during decoding,
such as changes in message values and tentative VN decisions,
to determine the order of message updates
\cite{RBP_NWRBP,SVNF,DSVNFS,RDRBP,UBPRBP,IVCRBP,OVRBP,V-VCRBP,TRM_RM}.
The residual belief-propagation (RBP) decoder
is an early IDS-based BP decoder for LDPC codes \cite{RBP_NWRBP}.
The RBP decoder defines the residual of a C2V message
as the magnitude of the difference
between its current and precomputed values.
At each update,
the RBP algorithm selects the C2V message with the largest residual,
propagates its precomputed value,
and then updates the related V2C messages.
Although the RBP decoder generally
provides favorable error-rate performance
in the first few decoding iterations,
the greedy message selection rule adopted
may cause a small group of edges
to be repeatedly chosen.
This behavior can waste decoding resources
by repeatedly updating the associated C2V and V2C messages,
thereby degrading the converged error-rate performance.
To address this greedy behavior,
several variants of the RBP decoder have been proposed.
The node-wise RBP (NWRBP) algorithm \cite{RBP_NWRBP}
follows the RBP decoder in identifying the C2V message
with the largest residual.
Instead of propagating only that message, however,
the NWRBP algorithm propagates all outgoing C2V messages
from the corresponding CN simultaneously,
thereby alleviating the greedy behavior
caused by the maximum-residual selection rule.
In \cite{SVNF},
two RBP variants were proposed
to prevent a small group of edges from dominating
the update process within an iteration:
the quota-based RBP algorithm limits the number of times
each edge can be updated;
the silent-variable-node-free RBP (SVNF-RBP) algorithm
processes the VNs in a fixed order
so that every VN can contribute its intrinsic information
to the decoding process.
In \cite{DSVNFS},
the dynamic SVNF-RBP (DSVNF-RBP) algorithm
dynamically determines the VN updating order
and gives higher updating priority
to C2V messages associated with unsatisfied CNs.
By scaling each residual with a factor that decreases
with the number of previous updates on the same edge,
the residual-decaying-based RBP (RDRBP) algorithm \cite{RDRBP}
makes edges that have been updated more frequently
less likely to be selected again within the same iteration.
From another perspective,
the update-before-propagate RBP (UBPRBP) algorithm \cite{UBPRBP}
considers whether a CN has received new V2C information
before propagating another C2V message from the same CN.
Once a CN has propagated a C2V message,
the CN is temporarily muted and cannot participate
in the subsequent message selection
until it receives a newly updated incoming V2C message.

In addition to modifying how C2V messages
are selected and propagated,
VN decision and reliability information
have also been used to improve RBP decoding
by determining which VN or message should be updated next
\cite{IVCRBP,OVRBP,V-VCRBP,TRM_RM}.
For example,
the informed variable-to-check residual BP (IVC-RBP) algorithm \cite{IVCRBP}
combines V2C message residual information
with the instability of a tentative VN decision;
the decision is regarded as unstable
when the current and precomputed total log-likelihood ratios (LLRs)
have different signs.
In \cite{OVRBP},
the oscillating-variable-node-based RBP (OVRBP) algorithm
was developed to evaluate the reliability of a VN
using its total LLR together with parity-check information;
the OVRBP algorithm also explicitly handles oscillating VNs.
The variable-node and variable-to-check-edge residual BP (V-VCRBP) algorithm
introduced in \cite{V-VCRBP}
combines a stability criterion
with residual measures for VNs and V2C messages.
In \cite{TRM_RM},
the reliability metric-based RBP (RM-RBP) algorithm
adopts a VN-wise updating procedure
that ensures every VN contributes its intrinsic information
and gives each edge an opportunity to be updated,
thereby alleviating greedy updating.
In the same work,
the TRM-TVRBP algorithm,
which is based on two reliability metrics
and two types of VN residuals,
was also proposed.
The TRM-TVRBP algorithm employs a two-step message selection strategy
that first determines the VN to be updated
according to VN reliability
and then selects the information used to update that VN.

With the aim of identifying and correcting erroneous or unreliable
tentative VN decisions as early as possible,
the work in \cite{CIRBP} adopted a VN-centric scheduling strategy
and introduced conditional innovation (CI)
as a metric for selecting the VN to be updated.
The CI metric is defined as the absolute difference
between the current and precomputed posterior probabilities.
This probability-domain formulation is motivated by the fact that
the posterior probability is directly related
to the reliability of a tentative bit decision,
while a change in the total LLR of a VN
does not translate linearly into
the corresponding change in its posterior probability
and therefore does not directly represent
the change in decision reliability.
The analysis in \cite{CIRBP} showed that
a larger CI indicates that the current tentative decision
is more likely to be erroneous and,
if erroneous,
more likely to be corrected after the VN is updated.
Based on this metric,
the CI-based RBP (CIRBP) decoder was developed
and was shown to provide significantly
improved error-rate performance.

Because the original CI metric depends only on the absolute difference
between the current and precomputed posterior probabilities,
different probability pairs having the same absolute difference
are assigned the same CI value.
The CI metric therefore cannot further distinguish among such pairs,
even though the individual posterior probabilities
may reflect different reliability levels
of their corresponding tentative decisions.
This limitation motivates us to further consider
the information conveyed by the individual posterior probabilities
when selecting the VN to be updated.
To characterize such reliability,
we use binary entropy,
which quantifies the uncertainty associated
with a binary posterior probability.
Accordingly,
we develop an entropy-based CI (E-CI) metric
that represents each posterior probability
by the complement of its binary entropy.
The E-CI metric compares the resulting representations
of the current and precomputed posterior probabilities
while retaining the decision-side information
associated with each probability.
Moreover,
because direct evaluation of the binary entropy
requires logarithmic operations,
we derive a simpler quadratic CI (Q-CI) metric
from the local quadratic behavior of the entropy complement
around the decision boundary,
thereby avoiding direct entropy evaluation.

We first evaluate the two proposed metrics
within the original CIRBP decoder
by replacing its CI metric with each proposed metric
while keeping all other decoding steps unchanged.
In this decoding procedure,
one incoming C2V message is selected
and propagated to the target VN at each update.
We also apply the two proposed metrics
to the node-wise extension of the CIRBP decoder,
in which all incoming C2V messages
are propagated to the selected VN within the same update.
Simulation results for the MacKay regular $(1008,504)$,
IEEE 802.11 $(1944,972)$,
and 5G NR $(3200,640)$ LDPC codes
show that both proposed metrics
can generally provide lower frame error rate (FER)
than that obtained with the original CI metric.
The Q-CI metric avoids direct entropy evaluation
and reduces the metric-evaluation complexity
associated with the E-CI metric
while maintaining favorable FER performance.
These results suggest that considering
the individual certainty levels
of the current and precomputed posterior probabilities
may provide additional information
for differentiating among VNs
when determining their update priorities,
beyond relying solely on the absolute difference
between the two posterior probabilities.

The rest of this paper is organized as follows.
Section~\ref{Sec2} provides the necessary preliminaries
and reviews the IDS-based BP decoders used for comparison,
including the CIRBP decoder and its CI metric.
Section~\ref{Sec3} develops the proposed E-CI and Q-CI metrics
as replacements for the original CI metric
and incorporates them into the generalized CIRBP decoder
and its node-wise extension.
Section~\ref{Sec4} presents the simulation results
and discusses the related complexity considerations.
Finally,
Section~\ref{Sec5} concludes this paper. 
\section{Preliminaries}
\label{Sec2}

\subsection{System Model and Conventional RBP Decoding}
\label{Sec2_A}

Consider a binary $(N,K)$ LDPC code $\mathcal{C}$,
where $N$ is the codeword length and $K$ is the number of information bits;
the code rate is $K/N$.
The code is characterized by an $M\times N$ parity-check matrix
$\bm{H}=[h_{mn}]$,
where $M$ is the number of parity-check equations
and $h_{mn}\in\{0,1\}$.
The corresponding Tanner graph contains $N$ VNs,
denoted by $v_n$, $0\leq n<N$,
and $M$ CNs,
denoted by $c_m$, $0\leq m<M$.
The VN $v_n$ corresponds to the $n$th coded bit
(i.e., the $n$th column of $\bm{H}$),
and the CN $c_m$ corresponds to the $m$th parity-check equation
(i.e., the $m$th row of $\bm{H}$).
An edge $(m,n)$ connects $c_m$ and $v_n$ if $h_{mn}=1$.
Let $E$ denote the total number of edges in the Tanner graph,
or equivalently, the number of nonzero entries in $\bm{H}$.

Let $\bm{u}=(u_0,u_1,\ldots,u_{N-1})$ denote a codeword,
where $u_n\in\{0,1\}$.
For binary phase-shift keying (BPSK) transmission
over an additive white Gaussian noise (AWGN) channel,
$u_n$ is mapped to $x_n=1-2u_n$,
and the received sample is given by $y_n=x_n+w_n$.
The noise samples $\{w_n\}_{n=0}^{N-1}$ are independent
and identically distributed (i.i.d.)
zero-mean Gaussian random variables
with variance $\sigma^2=N_0/2$,
where $N_0/2$ is the two-sided noise power spectral density.
The channel LLR associated with $v_n$ is
$L_n^{\mathrm{ch}}\triangleq 2y_n/\sigma^2$.

Given these channel LLRs,
the BP decoder performs 
iterative message updates on the Tanner graph.
To describe these updates,
let $\mathcal{M}(n)=\{m\mid h_{mn}=1\}$ and
$\mathcal{N}(m)=\{n\mid h_{mn}=1\}$
respectively denote the index sets of the CNs connected to VN $v_n$
and the VNs connected to CN $c_m$.
The corresponding sets with one index excluded are denoted by
$\mathcal{M}(n)\setminus\{m\}$ and
$\mathcal{N}(m)\setminus\{n\}$, respectively.
Let $L^{\mathrm{c2v}}_{(m,n)}$ denote the C2V message
from $c_m$ to $v_n$,
and let $L^{\mathrm{v2c}}_{(n,m)}$ denote the V2C message
from $v_n$ to $c_m$.
At the beginning of BP decoding,
the messages are initialized as
$L^{\mathrm{c2v}}_{(m,n)}=0$
and
$L^{\mathrm{v2c}}_{(n,m)}=L_n^{\mathrm{ch}}$
for all edges.
The BP decoder performs the C2V and V2C message updates according to
\begin{equation}
L^{\mathrm{c2v}}_{(m,n)}
=
2\tanh^{-1}\!\left[
\prod_{n'\in\mathcal{N}(m)\setminus\{n\}}
\tanh\!\left(
\frac{L^{\mathrm{v2c}}_{(n',m)}}{2}
\right)
\right],
\label{eq:C2V}
\end{equation}
and
\begin{equation}
L^{\mathrm{v2c}}_{(n,m)}
=
L_n^{\mathrm{ch}}
+
\sum_{m'\in\mathcal{M}(n)\setminus\{m\}}
L^{\mathrm{c2v}}_{(m',n)}.
\label{eq:V2C}
\end{equation}
The total LLR used to obtain the tentative decision at VN $v_n$ is
\begin{equation}
L_n
=
L_n^{\mathrm{ch}}
+
\sum_{m\in\mathcal{M}(n)}
L^{\mathrm{c2v}}_{(m,n)}.
\label{eq:totalLLR}
\end{equation}
Define the hard-decision function as
$\operatorname{HD}(x)=0$ for $x\geq0$
and $\operatorname{HD}(x)=1$ otherwise.
The tentative bit decision associated 
with VN $v_n$ is then obtained as
$\hat u_n=\operatorname{HD}(L_n)$,
and the resulting tentative decision vector is
$\hat{\bm u}=(\hat u_0,\ldots,\hat u_{N-1})$.
To verify whether $\hat{\bm u}$ is a valid codeword,
the syndrome vector $\bm{s}=(s_0,\ldots,s_{M-1})$ is computed as
$\bm{s}=\bm{H}\hat{\bm u}^{\mathrm T}\pmod{2}$,
and the tentative decision vector $\hat{\bm u}$ 
is a valid codeword
if and only if $\bm{s}=\bm{0}$.
The parity-check equation corresponding to CN $c_m$
is satisfied if $s_m=0$;
otherwise, $c_m$ is referred to as an unsatisfied CN.
The BP decoding process usually terminates
when all parity-check equations are satisfied
or when the number of decoding iterations reaches
the predetermined maximum, denoted by $I_{\max}$.

For each edge $(m,n)$,
the RBP algorithm computes the precomputed C2V message
$\widetilde{L}^{\mathrm{c2v}}_{(m,n)}$
using the currently available V2C messages
incoming to $c_m$ according to (\ref{eq:C2V}),
but does not propagate it immediately.
Instead, the RBP algorithm first computes
the corresponding C2V message residual as
\begin{equation}
R^{\mathrm{c2v}}_{(m,n)}
=
\left|
\widetilde{L}^{\mathrm{c2v}}_{(m,n)}
-
L^{\mathrm{c2v}}_{(m,n)}
\right|.
\label{eq:C2Vresidual}
\end{equation}
The RBP algorithm then selects the edge
with the largest residual according to
\begin{equation}
(m^*,n^*)
=
\arg\max_{(m,n):h_{mn}=1}
R^{\mathrm{c2v}}_{(m,n)}.
\label{eq:RBP_select}
\end{equation}
The precomputed C2V message on the selected edge $(m^*,n^*)$
then replaces the current C2V message
and is propagated to $v_{n^*}$,
followed by updates of the total LLR of $v_{n^*}$
and the associated outgoing V2C messages.
The CNs receiving the newly updated V2C messages
then recompute the precomputed C2V messages
and the corresponding residuals
before the next message selection.

\subsection{Related IDS-Based BP Decoders and the Conditional Innovation (CI) Metric}
\label{Sec2_B}

This subsection provides a brief review of the IDS-based BP decoders
used for comparison in Section~\ref{Sec4}.
The NWRBP decoder \cite{RBP_NWRBP}
identifies the edge with the largest C2V residual,
as in the RBP decoder,
but propagates all precomputed C2V messages
from the corresponding CN
rather than only the message on the selected edge.
The SVNF-RBP algorithm \cite{SVNF}
processes the VNs in the natural VN index order,
$v_0,v_1,\ldots,v_{N-1}$.
For each VN in this order,
the incoming C2V message with the largest residual is propagated,
and the updated VN then generates its outgoing V2C messages,
ensuring that the intrinsic information of every VN
can be used in the decoding process
and preventing any VN from remaining silent.
The DSVNF-RBP decoder \cite{DSVNFS}
relaxes the fixed VN-order constraint 
imposed by the SVNF-RBP algorithm.
It computes on-demand C2V residuals
associated with the most recently updated VN and CN,
gives priority to candidate edges connected to unsatisfied CNs,
and selects the candidate with the largest residual
for the next message update.
The RDRBP algorithm proposed in \cite{RDRBP}
attenuates the residual of an edge
according to the number of times
that edge has already been chosen to be updated 
in the current iteration.
Let $\nu_{(m,n)}$ denote this update count,
which is reset to zero at the start of each iteration.
The decayed residual is then given by
\begin{equation}
R^{\mathrm{RD}}_{(m,n)}
=
\beta^{\nu_{(m,n)}}
R^{\mathrm{c2v}}_{(m,n)},
\qquad
0<\beta\leq1,
\end{equation}
where $\beta$ is the residual-decay factor.
At each update,
the UBPRBP decoder \cite{UBPRBP}
compares the largest C2V residual
associated with each unmuted CN.
After a C2V message is propagated,
its source CN is muted,
so that all C2V residuals associated with this CN
are temporarily excluded from message selection.
These C2V residuals are allowed to participate
in the message selection again
only after the CN receives new V2C information
from one of its neighboring VNs.
The OVRBP decoder \cite{OVRBP}
adopts a VN-level updating strategy
and explicitly processes oscillating VNs.
Its dynamic selection rule ranks VNs
according to total LLR sign changes,
the number of adjacent unsatisfied CNs,
and VN residuals.
After selecting a target VN for update,
the OVRBP decoder updates all of its incoming C2V messages.
When these updates reverse the sign of the VN's total LLR,
its tentative bit decision also changes,
and the OVRBP decoder identifies this event as VN oscillation.
With the oscillating VNs marked,
the decoder then applies its oscillation-processing rule
before generating the outgoing V2C messages.

Changes in a VN's total LLR have been used
in several IDS-based BP decoders
to assess VN reliability
and hence to guide the selection of VNs or edges to be updated
\cite{IVCRBP,OVRBP,V-VCRBP,TRM_RM}.
The CIRBP algorithm also focuses on VN-level reliability,
but evaluates the current and precomputed VN information
in the probability domain
rather than directly using the change in total LLR.
For this purpose,
the precomputed total LLR of $v_n$ is first defined as
\begin{equation}
\widetilde{L}_n
=
L_n^{\mathrm{ch}}
+
\sum_{m\in\mathcal{M}(n)}
\widetilde{L}^{\mathrm{c2v}}_{(m,n)}.
\label{eq:precomputed_totalLLR}
\end{equation}
The current and precomputed total LLRs
are then mapped to their corresponding
conditional posterior probabilities \cite{CIRBP}.
For an arbitrary LLR value $\ell$,
these probabilities are defined as
\begin{equation}
p_0(\ell)
=
\frac{1}{1+e^{-\ell}},
\qquad
p_1(\ell)
=
1-p_0(\ell),
\label{eq:posterior_probability}
\end{equation}
where $p_0(\ell)$ and $p_1(\ell)$
are the posterior probabilities
that the coded bit takes the values $0$ and $1$, respectively.
The original CI metric for $v_n$ is therefore given by
\begin{equation}
F_n^{\mathrm{CI}}
=
\left|p_0(\widetilde{L}_n)-p_0(L_n)\right|
=
\left|p_1(\widetilde{L}_n)-p_1(L_n)\right|,
\qquad
0\leq F_n^{\mathrm{CI}}<1,
\label{eq:CI}
\end{equation}
where the second equality follows from
(\ref{eq:posterior_probability}).

The two properties established in \cite{CIRBP}
provide the basis for using CI
as a metric for selecting the VN to be updated.
Assuming that the C2V messages
can be modeled as i.i.d. Gaussian random variables
and that the signal-to-noise ratio (SNR) is sufficiently high,
a larger CI indicates that
the current tentative decision of a VN
is more likely to be erroneous and,
conditioned on an erroneous decision,
more likely to be corrected after an update.
Therefore,
the CIRBP decoder first identifies
the VN with the largest CI value
as the candidate VN for the next update.
However,
when the CI values of all VNs are small,
CI alone may be less effective
in identifying a VN whose tentative decision is erroneous
and likely to be corrected by an update.
To account for this case,
the CIRBP decoder introduces a threshold $\gamma_{\mathrm{CI}}$.
If the largest CI value is at least $\gamma_{\mathrm{CI}}$,
the candidate VN is selected as the target VN,
and the residual search is restricted
to its incoming C2V messages,
among which the message with the largest residual
is propagated.
Otherwise,
a global residual search is performed
over all edges in the Tanner graph,
and the C2V message with the largest residual
is selected for propagation,
which corresponds to the original RBP selection rule
\cite{RBP_NWRBP}.

\section{Proposed Enhanced CI Metrics and Decoding Algorithms}
\label{Sec3}

\subsection{Reformulation of the Original CI Metric}
\label{Sec3_A}

The original CI metric in (\ref{eq:CI}) is determined
by the absolute difference 
between the current and precomputed posterior probabilities.
For convenience, denote these two probabilities by
$p_n\triangleq p_0(L_n)$ and
$\widetilde p_n\triangleq p_0(\widetilde L_n)$, respectively, 
where $p_n$ and $\widetilde p_n$ 
are the current and precomputed posterior probabilities 
that the coded bit associated with $v_n$ is $0$.
The original CI metric can then be written as
\begin{equation}
F_n^{\mathrm{CI}}
=
|\widetilde p_n-p_n|.
\label{eq:CI_restatement}
\end{equation}

To further examine the information represented
by each posterior probability in (\ref{eq:CI_restatement}),
we now express it relative to the decision boundary $p=0.5$.
Define the sign function as
$\operatorname{sgn}(x)=1$, $0$, and $-1$ for
$x>0$, $x=0$, and $x<0$, respectively.
For any binary posterior probability $p\in[0,1]$,
\[
p-0.5
=
|p-0.5|\operatorname{sgn}(p-0.5).
\]
Since
$\widetilde p_n-p_n
=(\widetilde p_n-0.5)-(p_n-0.5)$,
the original CI metric can be equivalently reformulated as
\begin{align}
F_n^{\mathrm{CI}}
=
\Big|
&|\widetilde p_n-0.5|
 \operatorname{sgn}(\widetilde p_n-0.5)
\nonumber\\
&-|p_n-0.5|
 \operatorname{sgn}(p_n-0.5)
\Big|.
\label{eq:CI_reformulated}
\end{align}

Since $p_0(0)=0.5$, the posterior probability $p=0.5$
corresponds to an LLR of zero
and represents the decision boundary 
between the two coded bit values.
In the representation in (\ref{eq:CI_reformulated}),
each posterior probability is described through two components:
$\operatorname{sgn}(p-0.5)$ indicates its side of the decision boundary,
and $|p-0.5|$ gives its linear distance from that boundary.
When $p$ is close to $0.5$, 
the two coded bit values
have similar posterior probabilities,
and the corresponding tentative decision is less reliable.
As $p$ moves away from $0.5$, 
the difference between the posterior probabilities
of the two coded bit values increases,
indicating a more reliable tentative decision.

Although the reformulation in (\ref{eq:CI_reformulated})
leaves the CI value unchanged, 
it separates the decision-side information
from the reliability information 
represented by each posterior probability.
In particular, the sign term identifies the side of the decision boundary,
and the magnitude term $|p-0.5|$ provides 
a measure of the reliability of the corresponding tentative decision.
This observation motivates replacing 
the linear-distance term $|p-0.5|$
with an alternative representation
that more directly characterizes the reliability
associated with each posterior probability.

\subsection{Proposed Entropy-Based CI (E-CI) and Quadratic CI (Q-CI) Metrics}
\label{Sec3_B}

To relate this reliability to an 
information-theoretic quantity,
we consider binary entropy,
which provides a natural measure of posterior uncertainty.
For a binary posterior probability $p$,
it is defined as
\begin{equation}
H(p)
=
-p\log_2 p-(1-p)\log_2(1-p),
\qquad 0\leq p\leq1,
\label{eq:entropy}
\end{equation}
with the convention $0\log_2 0=0$.
Since a larger $H(p)$ indicates greater posterior uncertainty
and hence a less reliable tentative decision,
we use the entropy complement $1-H(p)$ 
to represent posterior certainty.
Note that although binary entropy 
and its complement are
defined over the closed interval $[0,1]$,
the posterior probabilities considered here 
are obtained from finite LLRs
according to (\ref{eq:posterior_probability})
and therefore satisfy $0<p<1$.

At the decision boundary $p=0.5$,
the coded bit values $0$ and $1$ 
have equal posterior probabilities,
and $1-H(p)=0$.
As $p$ moves toward either $0$ or $1$,
the posterior probability of one coded bit value increases
while that of the other decreases,
and $1-H(p)$ increases toward $1$.
Thus, $1-H(p)$ and the linear distance $|p-0.5|$
exhibit the same monotonic trend:
both are zero at the decision boundary
and increase as the posterior probability moves toward either $0$ or $1$.
The entropy complement therefore provides
an entropy-based representation of posterior certainty
that varies consistently with the decision reliability
represented by the linear-distance term
in the reformulated original CI metric.
Because $H(p)=H(1-p)$,
posterior probabilities symmetrically located with respect to $0.5$
have the same entropy-complement value.
This symmetry is reasonable from the perspective of decision reliability,
because posterior probabilities approaching $0$ and $1$
correspond to equally reliable tentative decisions,
although they favor opposite coded bit values.
However, the entropy complement alone cannot distinguish
on which side of the decision boundary the posterior probability lies.
We therefore retain the factor $\operatorname{sgn}(p-0.5)$
to preserve the decision-side information represented in
(\ref{eq:CI_reformulated}).

Based on these properties,
the entropy-based CI (E-CI) metric is defined as
\begin{align}
F_n^{\mathrm{E\text{-}CI}}
\triangleq
\Big|
&\operatorname{sgn}(\widetilde p_n-0.5)\,[1-H(\widetilde p_n)]
\nonumber\\
&-\operatorname{sgn}(p_n-0.5)\,[1-H(p_n)]
\Big|,
\qquad 0\leq F_n^{\mathrm{E\text{-}CI}}<2.
\label{eq:E-CI}
\end{align}
For the current and precomputed posterior probabilities,
the E-CI metric first forms their respective signed entropy complements.
Here, $1-H(p)$ represents posterior certainty,
whereas $\operatorname{sgn}(p-0.5)$ indicates
the side of the decision boundary
on which the posterior probability lies.
The E-CI metric then takes the absolute difference
between the two signed entropy complements.
Compared with the reformulated CI metric in (\ref{eq:CI_reformulated}),
the proposed E-CI metric retains the sign terms
that distinguish the decision sides
and the absolute-difference structure used to compare
the two signed representations,
but replaces each linear-distance term $|p-0.5|$
with the corresponding entropy complement $1-H(p)$.

Relative to the original CI metric,
direct evaluation of the proposed E-CI metric
incurs additional computational cost
because the computation of binary entropy
requires logarithmic operations.
To address this computational overhead,
we approximate the entropy complement
in a neighborhood of the decision boundary $p=0.5$
and use this approximation to obtain a simpler metric.
Let $p=0.5+\delta$, where
$\delta=p-0.5\in(-0.5,0.5)$ denotes the signed deviation
of $p$ from the decision boundary for a finite LLR.
The expansion of $1-H(p)$ about $p=0.5$
(equivalently, $\delta=0$) is
\begin{equation}
1-H(0.5+\delta)
=
\frac{2}{\ln 2}\delta^2
+
\frac{4}{3\ln 2}\delta^4
+
O(\delta^6).
\label{eq:entropy_expansion}
\end{equation}
As shown in (\ref{eq:entropy_expansion}),
the constant and linear terms are both zero.
Therefore, when $|\delta|$ is sufficiently small,
the quadratic term is the leading nonzero term.
As a result, in the vicinity of the decision boundary,
the entropy complement can be approximated
by keeping only this quadratic term.
Substitution of this approximation into (\ref{eq:E-CI}) yields
\begin{align}
F_n^{\mathrm{E\text{-}CI}}
\approx
\frac{2}{\ln 2}
\Big|
&\operatorname{sgn}(\widetilde p_n-0.5)(\widetilde p_n-0.5)^2
\nonumber\\
&-\operatorname{sgn}(p_n-0.5)(p_n-0.5)^2
\Big|.
\label{eq:E-CI_local}
\end{align}
Moreover, since the positive factor $2/\ln 2$
in (\ref{eq:E-CI_local})
does not affect the ordering of the metric values among VNs,
we omit this common factor
and define the quadratic CI (Q-CI) metric as
\begin{align}
F_n^{\mathrm{Q\text{-}CI}}
\triangleq
\Big|
&\operatorname{sgn}(\widetilde p_n-0.5)(\widetilde p_n-0.5)^2
\nonumber\\
&-\operatorname{sgn}(p_n-0.5)(p_n-0.5)^2
\Big|,
\qquad 0\leq F_n^{\mathrm{Q\text{-}CI}}<0.5.
\label{eq:Q-CI}
\end{align}
Note that the strict upper bounds
in (\ref{eq:E-CI}) and (\ref{eq:Q-CI})
follow from the fact that the posterior probabilities
obtained from finite LLRs
satisfy $0<p<1$.
The corresponding limiting values can only be approached
as one of the relevant LLRs tends to $+\infty$
and the other tends to $-\infty$.

The squared-distance term $(p-0.5)^2$ used 
in the Q-CI metric
follows from the leading nonzero term
in the expansion of the entropy complement 
about the decision boundary.
Hence, when the current and precomputed posterior probabilities
are both sufficiently close to the decision boundary,
the E-CI and Q-CI metrics satisfy
\begin{align}
F_n^{\mathrm{E\text{-}CI}}
\approx
\frac{2}{\ln 2}F_n^{\mathrm{Q\text{-}CI}}.
\end{align}
This local proportionality does not generally hold
over the entire probability interval.
Nevertheless, the local expansion provides a theoretical basis
for replacing the linear-distance term in the original CI metric
with a squared-distance term in the Q-CI metric.
In particular, the linear distance $|p-0.5|$
used in the original CI metric
provides a monotonic indication of decision reliability,
but does not capture the quadratic leading-order behavior
of the entropy complement near the decision boundary.
This neighborhood is particularly relevant
to determining the update priorities of VNs,
since posterior probabilities close to $0.5$ correspond to
high posterior uncertainty and hence less reliable tentative decisions.
In contrast, the squared distance $(p-0.5)^2$
matches the leading-order variation of $1-H(p)$ in this region
and therefore provides a quadratic representation
of the entropy-based posterior certainty.
Viewed from the reformulated original CI metric
in (\ref{eq:CI_reformulated}),
the Q-CI metric preserves the same decision-side information
and absolute-difference structure,
while replacing the linear distance $|p-0.5|$
with the squared distance $(p-0.5)^2$.
Thus, the Q-CI metric can be viewed as an entropy-motivated
quadratic modification of the original CI metric
that is locally proportional to the E-CI metric,
with the proportionality factor given by $2/\ln 2$,
while avoiding the logarithmic operations
required for direct evaluation of the E-CI metric.

\subsection{Proposed E-CIRBP and Q-CIRBP Decoders}
\label{Sec3_C}

Replacing the CI metric with the proposed E-CI or Q-CI metric
in the VN-selection step of the original CIRBP algorithm
yields the E-CIRBP and Q-CIRBP decoders, respectively.
For convenience,
the CIRBP, E-CIRBP, and Q-CIRBP decoders are described
under a unified notation:
$X\in\{\mathrm{CI},\mathrm{E\text{-}CI},\mathrm{Q\text{-}CI}\}$
denotes the adopted metric,
and $F_n^X$ denotes the corresponding metric value for $v_n$.
For each metric $X$,
an associated threshold $\gamma_X$
is used to determine whether the candidate VN
identified by the metric is selected as the target VN
for the current update.
If the largest metric value is below $\gamma_X$,
the decoder performs the global maximum-residual edge search
used in the original RBP algorithm.
Otherwise,
the candidate VN is selected as the target VN,
and the maximum-residual search is restricted
to its incoming edges.
The thresholds associated with
the CI, E-CI, and Q-CI metrics
are denoted by
$\gamma_{\mathrm{CI}}$, $\gamma_{\mathrm{E\text{-}CI}}$, and $\gamma_{\mathrm{Q\text{-}CI}}$, respectively.
The resulting general decoding procedure is summarized
in Algorithm~\ref{alg:CIRBP_general}.

\begin{algorithm}[h]
\caption{General CI-Based RBP Decoder}
\label{alg:CIRBP_general}
\begin{algorithmic}[1]
\STATE Initialize $L^{\mathrm{c2v}}_{(m,n)}=0$ for every edge
\STATE Set $L^{\mathrm{v2c}}_{(n,m)}=L_n^{\mathrm{ch}}$ for every edge
and $L_n=L_n^{\mathrm{ch}}$ for every VN
\STATE Generate all $\widetilde{L}^{\mathrm{c2v}}_{(m,n)}$
and compute all $R^{\mathrm{c2v}}_{(m,n)}$
\STATE Compute all $\widetilde{L}_n$ and $F_n^X$
\REPEAT
\STATE Find $n^*=\arg\max_{0\leq n<N}F_n^X$ \label{line:CIRBP:select_vn}
\IF{$F_{n^*}^X<\gamma_X$} \label{line:CIRBP:threshold}
\STATE Find $(m^*,n^*)=\arg\max_{(m,n):h_{mn}=1}R^{\mathrm{c2v}}_{(m,n)}$ \label{line:CIRBP:global_search}
\ELSE
\STATE Find $m^*=\arg\max_{m\in\mathcal{M}(n^*)}R^{\mathrm{c2v}}_{(m,n^*)}$ \label{line:CIRBP:local_search}
\ENDIF
\STATE Set $L^{\mathrm{c2v}}_{(m^*,n^*)}
\leftarrow\widetilde{L}^{\mathrm{c2v}}_{(m^*,n^*)}$
and propagate it to $v_{n^*}$ \label{line:CIRBP:propagate_c2v}
\STATE Set $R^{\mathrm{c2v}}_{(m^*,n^*)}=0$
and update $L_{n^*}$ and $F_{n^*}^X$ \label{line:CIRBP:update_target}
\FOR{every $m'\in\mathcal{M}(n^*)\setminus\{m^*\}$}
\STATE Generate and propagate $L^{\mathrm{v2c}}_{(n^*,m')}$ \label{line:CIRBP:propagate_v2c}
\FOR{every $n'\in\mathcal{N}(m')\setminus\{n^*\}$}
\STATE Recompute
$\widetilde{L}^{\mathrm{c2v}}_{(m',n')}$,
$R^{\mathrm{c2v}}_{(m',n')}$,
$\widetilde{L}_{n'}$, and $F_{n'}^X$ \label{line:CIRBP:recompute_neighbor}
\ENDFOR
\ENDFOR
\UNTIL{the stopping condition is satisfied}
\end{algorithmic}
\end{algorithm}

At the initialization of Algorithm~\ref{alg:CIRBP_general},
all current C2V messages are set to zero,
while every V2C message and VN total LLR is initialized
to the channel LLR of the associated VN.
The decoder then generates all precomputed C2V messages
and computes their residuals.
For each VN,
it also computes the precomputed total LLR
$\widetilde{L}_n$ and the metric value $F_n^X$,
where $F_n^X$ is evaluated using
the CI, E-CI, or Q-CI metric,
as specified by $X$.
At the beginning of each update,
the decoder identifies the VN associated with the largest metric value by
$n^*=\arg\max_{0\leq n<N}F_n^X$.
This VN serves as the candidate VN for the current update.
The subsequent threshold test determines
whether $v_{n^*}$ is selected as the target VN
or whether the global maximum-residual edge search is performed
(lines~\ref{line:CIRBP:select_vn}--\ref{line:CIRBP:local_search}).
If $F_{n^*}^X<\gamma_X$,
the decoder performs the global maximum-residual edge search
used in the original RBP algorithm.
That is,
the global search jointly determines the VN to be updated
and its incoming edge,
and may therefore replace
the candidate VN initially identified by the metric.
If $F_{n^*}^X\geq\gamma_X$,
the candidate VN is selected as the target VN,
and the maximum-residual search is restricted
to its incoming edges.

After $(m^*,n^*)$ has been determined in either case,
the corresponding precomputed C2V message
$\widetilde{L}^{\mathrm{c2v}}_{(m^*,n^*)}$
replaces the current C2V message
and is propagated to $v_{n^*}$
(line~\ref{line:CIRBP:propagate_c2v}).
The decoder then sets the residual on the selected edge to zero
and updates $L_{n^*}$ and $F_{n^*}^X$.
Next,
$v_{n^*}$ generates and propagates
$L^{\mathrm{v2c}}_{(n^*,m')}$
to every adjacent CN $c_{m'}$ other than $c_{m^*}$.
For the other VNs connected to these CNs,
the decoder recomputes the corresponding precomputed C2V messages,
residuals, precomputed total LLRs, and metric values
before the next update
(lines~\ref{line:CIRBP:propagate_v2c}--\ref{line:CIRBP:recompute_neighbor}).

Thus,
the E-CIRBP and Q-CIRBP decoders differ from the CIRBP decoder
only in the metric used to identify the candidate VN
and its associated threshold;
all other decoding steps remain unchanged.

\subsection{Proposed NW-E-CIRBP and NW-Q-CIRBP Decoders}
\label{Sec3_D}

Algorithm~\ref{alg:CIRBP_general}
propagates only one C2V message at each update.
We further extend this decoding operation
to a VN-centric node-wise form,
in which multiple C2V messages associated with the selected VN
are propagated within the same update.
This operation is similar to the multi-message propagation used in
the NWRBP decoder \cite{RBP_NWRBP},
but here the propagated messages are determined by the selected VN.  
Specifically,
the VN with the largest $F_n^X$
is directly selected as the target VN for updating.
All of its incoming precomputed C2V messages
then replace the corresponding current C2V messages
and are propagated to the selected VN within the same update.
Since the target VN is selected directly according to $F_n^X$
and all of its incoming C2V messages are updated,
neither a threshold comparison
nor a residual-based edge search is required.
The decoders employing
the CI, E-CI, and Q-CI metrics
under this node-wise (NW) operation are denoted by
NW-CIRBP, NW-E-CIRBP, and NW-Q-CIRBP, respectively.
Their common decoding procedure is summarized
in Algorithm~\ref{alg:NW_CIRBP_general}.

\begin{algorithm}[h]
\caption{General VN-Centric Node-Wise CI-Based RBP Decoder}
\label{alg:NW_CIRBP_general}
\begin{algorithmic}[1]
\STATE Initialize $L^{\mathrm{c2v}}_{(m,n)}=0$ for every edge
\STATE Set $L^{\mathrm{v2c}}_{(n,m)}=L_n^{\mathrm{ch}}$ for every edge
and $L_n=L_n^{\mathrm{ch}}$ for every VN
\STATE Generate all $\widetilde{L}^{\mathrm{c2v}}_{(m,n)}$
\STATE Compute all $\widetilde{L}_n$ and $F_n^X$
\REPEAT
\STATE Find $n^*=\arg\max_{0\leq n<N}F_n^X$ \label{line:NW:select_vn}
\FOR{every $m\in\mathcal{M}(n^*)$}
\STATE Set $L^{\mathrm{c2v}}_{(m,n^*)}
\leftarrow\widetilde{L}^{\mathrm{c2v}}_{(m,n^*)}$
and propagate it to $v_{n^*}$ \label{line:NW:propagate_c2v}
\ENDFOR
\STATE Set $L_{n^*}=\widetilde L_{n^*}$
and $F_{n^*}^X=0$ \label{line:NW:update_target}
\FOR{every $m'\in\mathcal{M}(n^*)$}
\STATE Generate and propagate $L^{\mathrm{v2c}}_{(n^*,m')}$ \label{line:NW:propagate_v2c}
\FOR{every $n'\in\mathcal{N}(m')\setminus\{n^*\}$}
\STATE Recompute
$\widetilde{L}^{\mathrm{c2v}}_{(m',n')}$,
$\widetilde{L}_{n'}$, and $F_{n'}^X$ \label{line:NW:recompute_neighbor}
\ENDFOR
\ENDFOR
\UNTIL{the stopping condition is satisfied}
\end{algorithmic}
\end{algorithm}

Algorithm~\ref{alg:NW_CIRBP_general}
uses the same initialization
for the C2V and V2C messages and VN total LLRs
as Algorithm~\ref{alg:CIRBP_general}.
At each update,
the decoder selects the VN $v_{n^*}$
with the largest metric value $F_n^X$
as the target VN (line~\ref{line:NW:select_vn}).
All incoming precomputed C2V messages of the selected VN
then replace their corresponding current C2V messages
and are propagated to $v_{n^*}$
(line~\ref{line:NW:propagate_c2v}).
After these messages have been propagated,
the decoder sets
$L_{n^*}=\widetilde L_{n^*}$ and $F_{n^*}^X=0$.
Using the updated total LLR,
$v_{n^*}$ then generates and propagates
its outgoing V2C messages to all adjacent CNs.
After each V2C message is propagated to an adjacent CN,
the decoder recomputes,
for every other VN connected to that CN,
the corresponding precomputed C2V message,
precomputed total LLR, and metric value
(lines~\ref{line:NW:propagate_v2c}--\ref{line:NW:recompute_neighbor}).
Once these updates have been completed for all adjacent CNs,
the decoder proceeds to the next VN selection.

Unlike Algorithm~\ref{alg:CIRBP_general},
Algorithm~\ref{alg:NW_CIRBP_general}
does not perform residual-based edge selection,
so C2V message residuals are not required.
Thus, both the exhaustive residual computation
at the beginning of decoding
and the residual updates associated with regenerated
precomputed C2V messages can be omitted. 
Moreover,
when the precomputed C2V messages are regenerated during decoding,
their corresponding residuals also do not need to be updated.
For Algorithms~\ref{alg:CIRBP_general}
and \ref{alg:NW_CIRBP_general},
one decoding iteration is defined to be completed
after a total of $E$ C2V messages have been propagated.
\section{Numerical Results}
\label{Sec4}

We evaluate the FER performance
obtained with the proposed E-CI and Q-CI metrics
under the two decoding procedures described in Section~\ref{Sec3}.
As described in Section~\ref{Sec2_A},
the simulations consider BPSK transmission
over an AWGN channel with $\sigma^2=N_0/2$.
Three LDPC codes are considered:
the MacKay regular $(1008,504)$ code \cite{MacKayWeb},
denoted by M-1008,
the IEEE 802.11 $(1944,972)$ LDPC code \cite{WiFi},
denoted by W-1944,
and the rate-$1/5$ $(3200,640)$ 5G NR LDPC code \cite{NR},
denoted by N-3200.
The N-3200 code is obtained by puncturing the first 128 VNs
of a length-3328 mother code generated from Base Graph 2 (BG2)
with lifting size $Z=64$.
These punctured VNs correspond to the first two block columns of BG2
and are not transmitted.
For the N-3200 code,
the metric evaluations, precomputed total-LLR calculations,
and VN updates are restricted to the core VNs.
The degree-one VNs in the extension part are not dynamically updated;
their channel LLRs are used as fixed V2C inputs
when generating C2V messages for the connected CNs
\cite{LPHD_RB-LBP,UBPRBP}.
The stopping condition is evaluated using
the parity-check equations associated with the core part.
For this code,
one decoding iteration is defined as the propagation of
a number of C2V messages equal to the total number of incoming edges
of the core-part VNs.
The same treatment of the core and extension parts
is applied to all considered decoding algorithms,
including the reference decoders, for a fair comparison.

The proposed E-CIRBP and Q-CIRBP decoders
are implemented according to Algorithm~\ref{alg:CIRBP_general},
and the reference decoders considered for the edge-wise updating case
include the RBP \cite{RBP_NWRBP},
SVNF-RBP \cite{SVNF},
DSVNF-RBP \cite{DSVNFS},
RDRBP \cite{RDRBP},
UBPRBP \cite{UBPRBP},
and the original CIRBP \cite{CIRBP} algorithms.
Similarly, the proposed NW-E-CIRBP and NW-Q-CIRBP decoders
are implemented according to Algorithm~\ref{alg:NW_CIRBP_general}.
The reference decoders considered for this case include
the NWRBP \cite{RBP_NWRBP} and OVRBP \cite{OVRBP} algorithms,
both of which employ node-wise IDS updating strategies
and propagate multiple C2V messages at each update,
as well as
the local-girth based grouped SBP (LG-GSBP) \cite{LG-GSBP}
and adaptive grouped SBP (AGSBP) \cite{AGSBP} algorithms,
which provide additional comparisons with SBP-based schedules
designed from the perspectives of
VN-oriented grouping and updating.
The parameter settings for the reference decoders,
including $\beta$ for RDRBP,
$\eta$ for AGSBP,
and the number of groups for LG-GSBP decoders,
are specified in the corresponding performance figures,
with the optimized values reported where applicable.

Let $I$ denote the number of decoding iterations performed
and $I_{\max}$ the maximum number of iterations
allowed in a simulation run.
For Algorithms~\ref{alg:CIRBP_general} and \ref{alg:NW_CIRBP_general},
one decoding iteration is completed
after a total of
$E$ C2V messages have been propagated,
following the definition given in Section~\ref{Sec3_D}.
For NWRBP, OVRBP, LG-GSBP, and AGSBP algorithms,
the iteration counts follow
their respective original definitions
\cite{RBP_NWRBP,OVRBP,LG-GSBP,AGSBP}.

\subsection{FER Performance and Convergence Behavior}
\label{Sec4_A}

\begin{figure}[!htbp]
    \centering
    \subfloat[M-1008 code; $E_b/N_0=2.75$ dB.]
    {\includegraphics[width=3.5in]{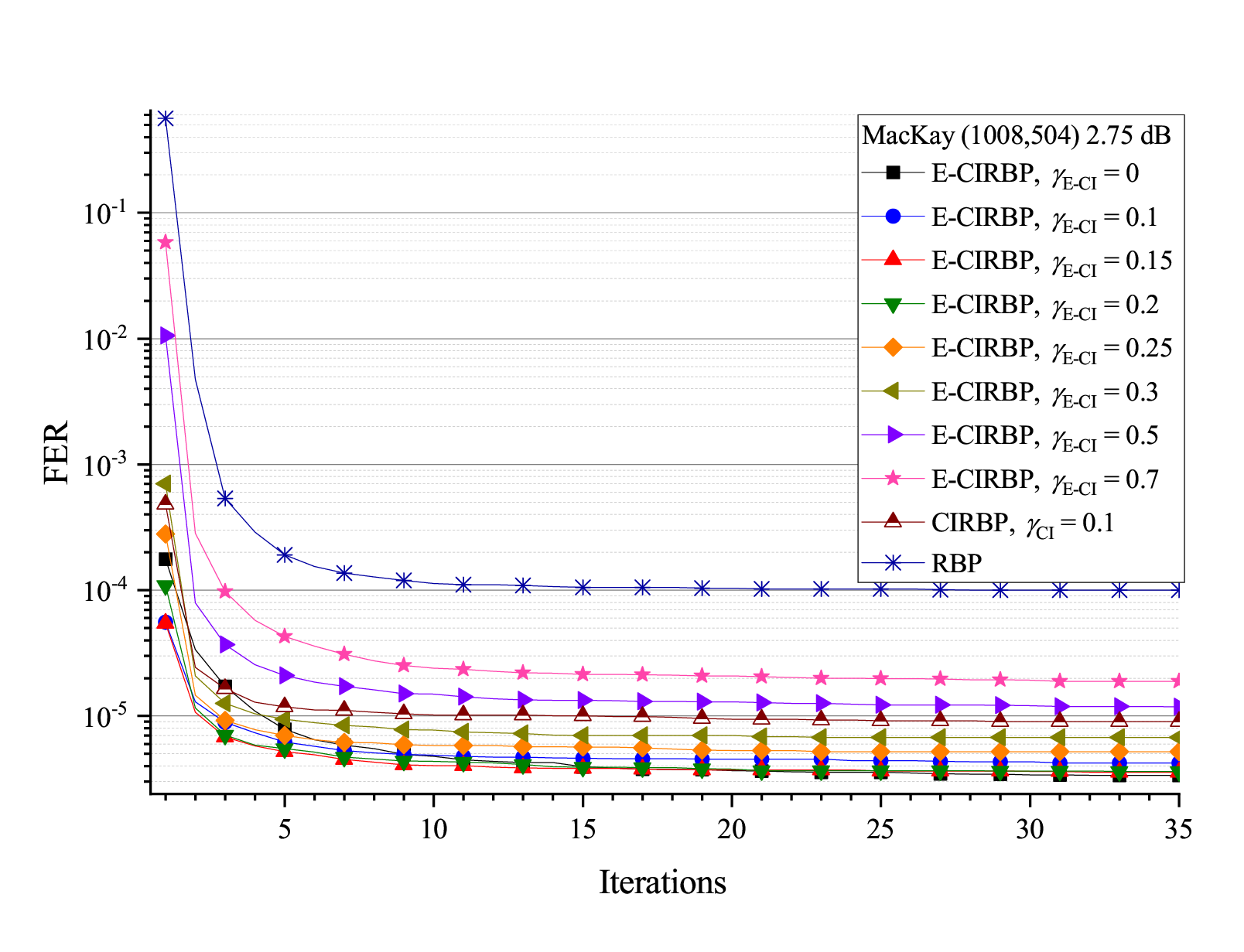}%
    \label{ITEvsFER_MK1008_EWIDS_ECI_gamma}}
    \hspace{0.00\textwidth}
    \subfloat[W-1944 code; $E_b/N_0=1.75$ dB.]
    {\includegraphics[width=3.5in]{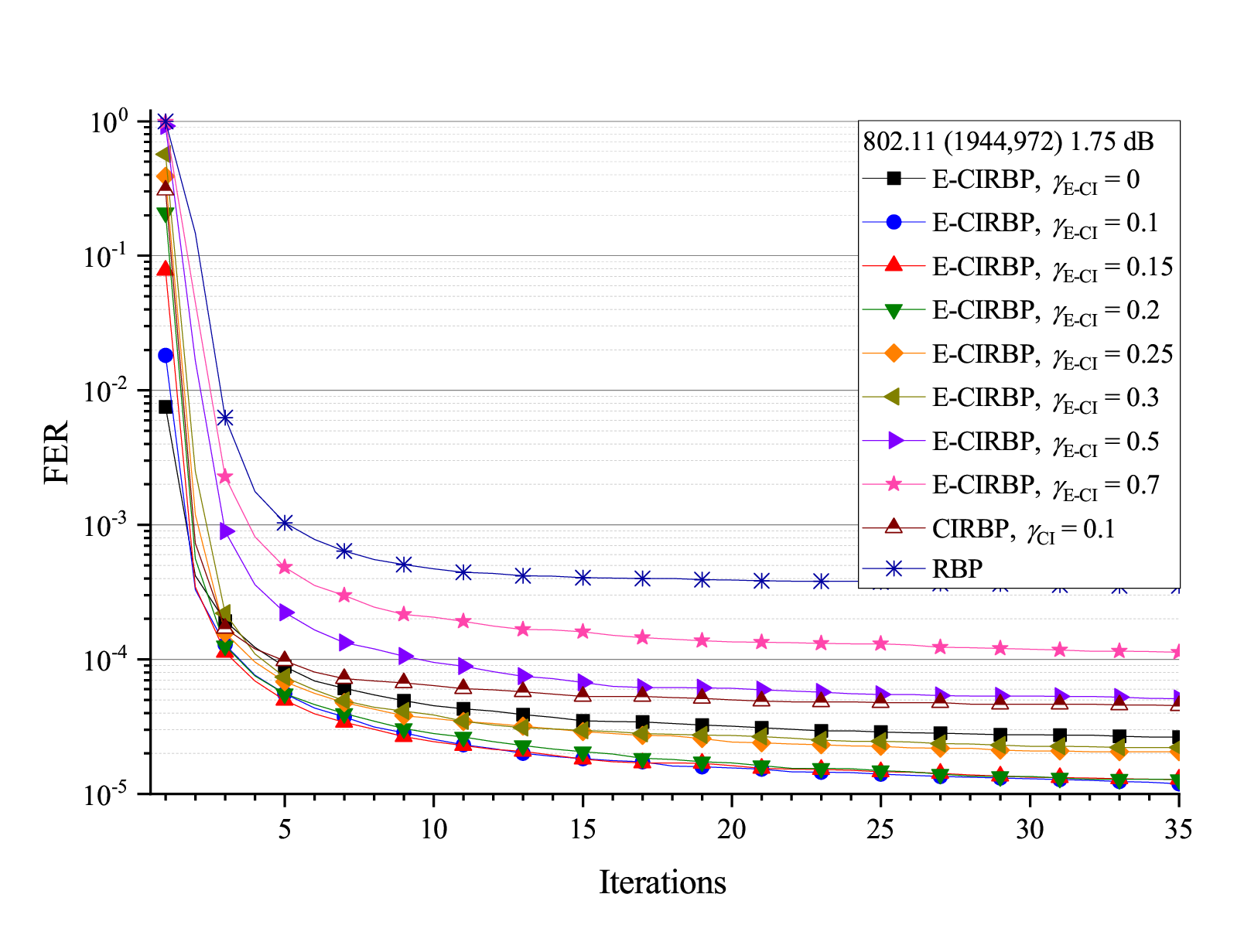}%
    \label{ITEvsFER_WIFI1944_EWIDS_ECI_gamma}}
    \hspace{0.00\textwidth}
    \subfloat[N-3200 code; $E_b/N_0=0.5$ dB.]
    {\includegraphics[width=3.5in]{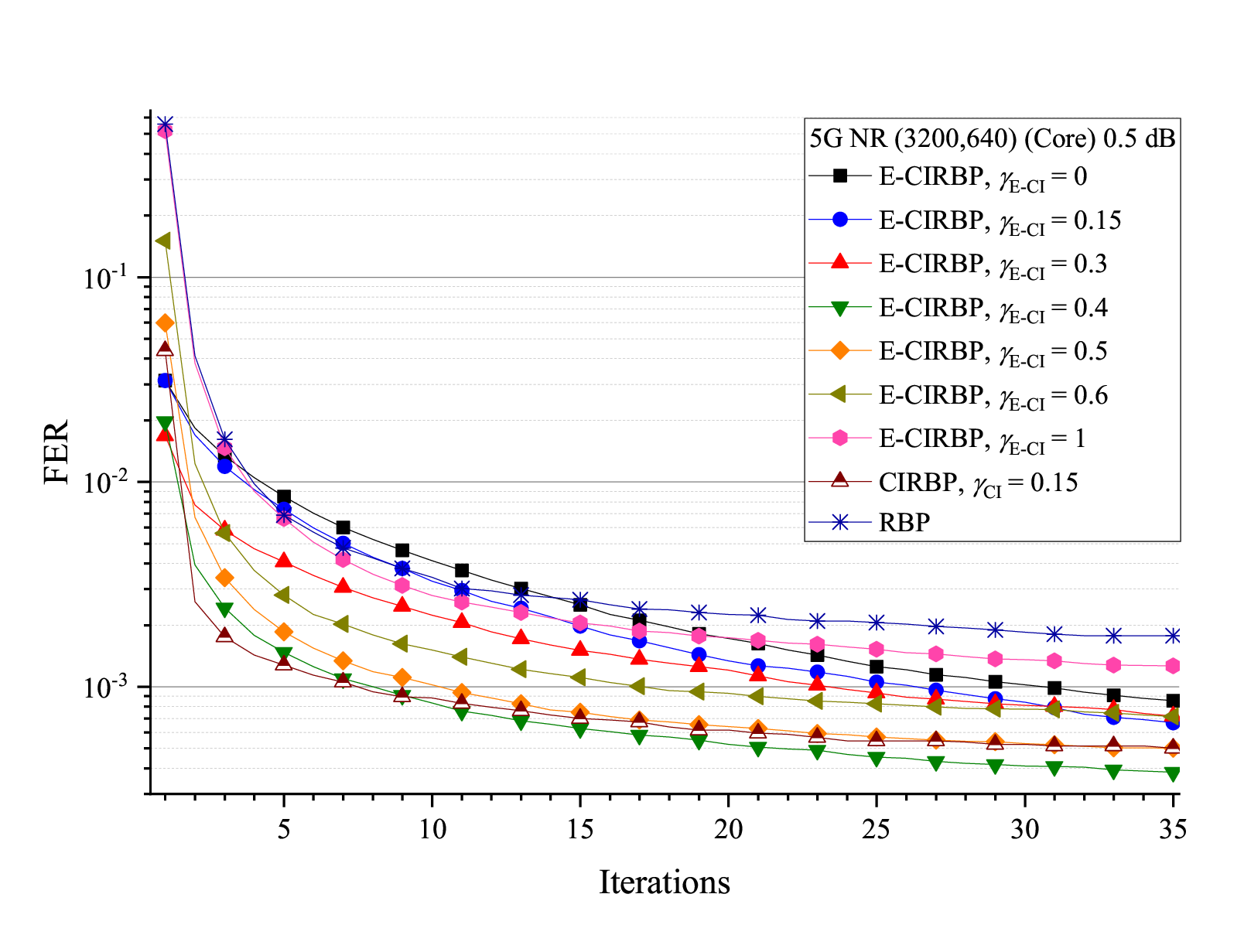}%
    \label{ITEvsFER_NR3200_EWIDS_ECI_gamma}}
    \caption{FER convergence behavior of the proposed E-CIRBP decoder
    for different $\gamma_{\mathrm{E\text{-}CI}}$ values
    with the rate-$1/2$ M-1008, rate-$1/2$ W-1944, and rate-$1/5$ N-3200 codes.}
    \label{Fig:ECI_gamma}
\end{figure}

\begin{figure}[!htbp]
    \centering
    \subfloat[M-1008 code; $E_b/N_0=2.75$ dB.]
    {\includegraphics[width=3.5in]{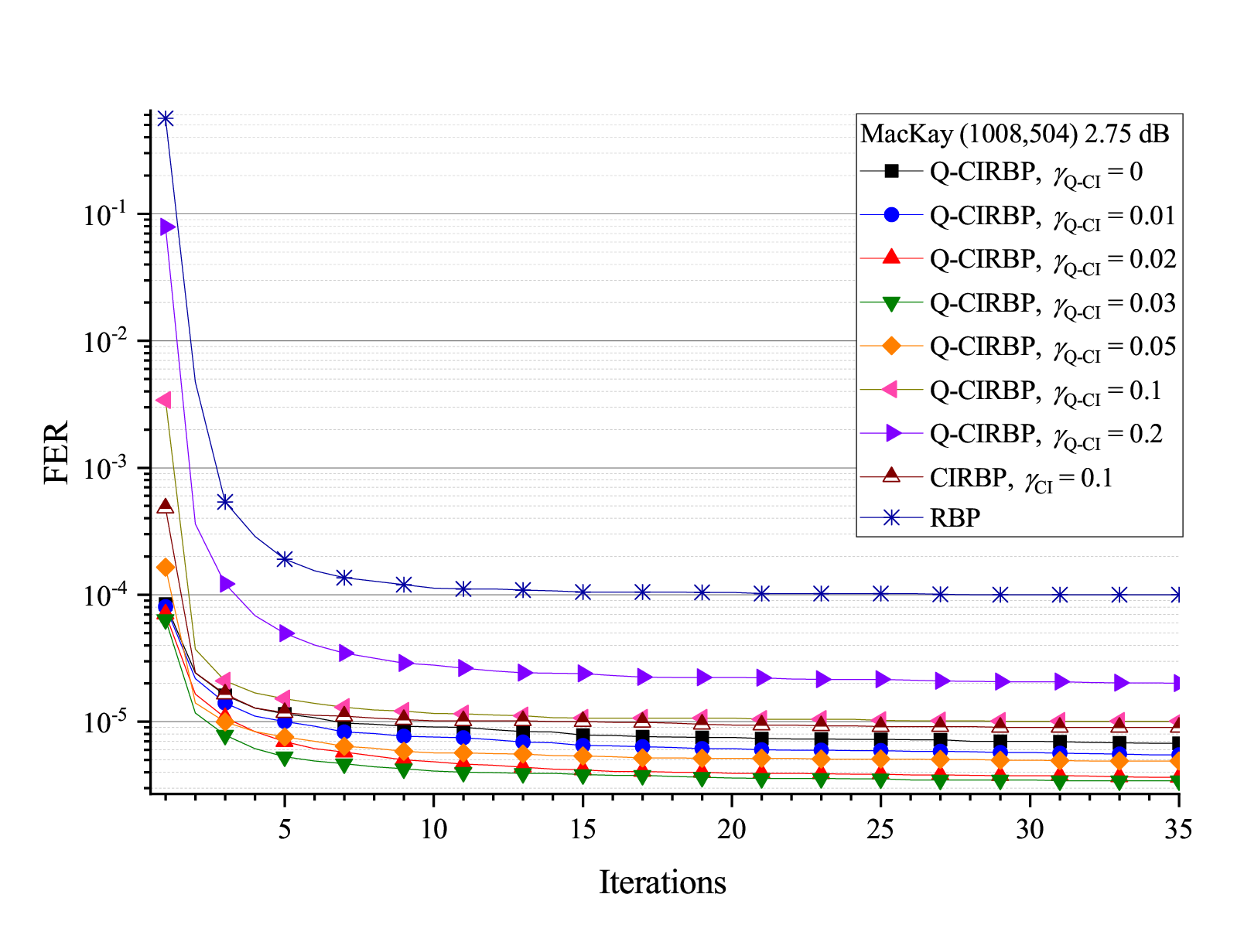}%
    \label{ITEvsFER_MK1008_EWIDS_QCI_gamma}}
    \hspace{0.00\textwidth}
    \subfloat[W-1944 code; $E_b/N_0=1.75$ dB.]
    {\includegraphics[width=3.5in]{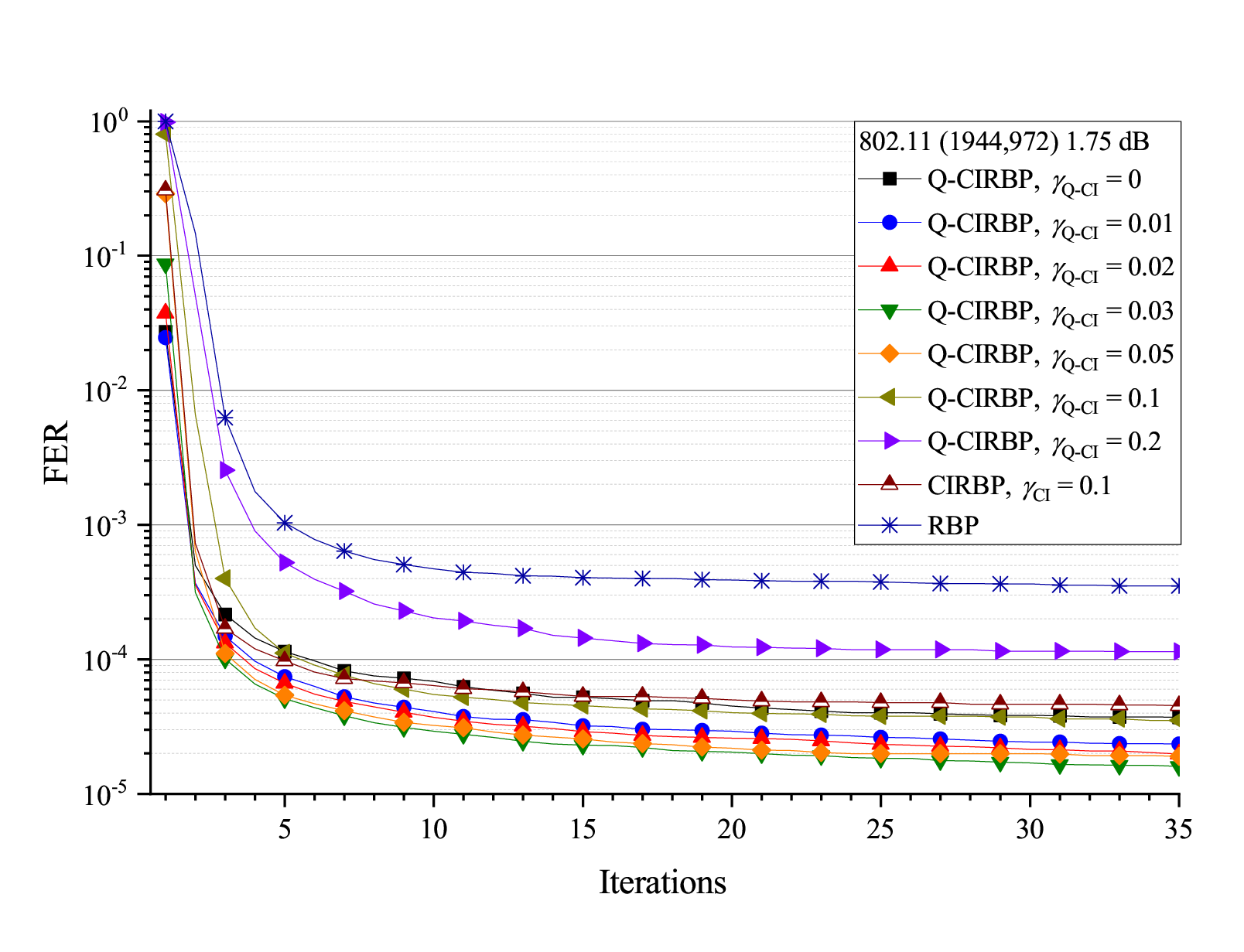}%
    \label{ITEvsFER_WIFI1944_EWIDS_QCI_gamma}}
    \hspace{0.00\textwidth}
    \subfloat[N-3200 code; $E_b/N_0=0.5$ dB.]
    {\includegraphics[width=3.5in]{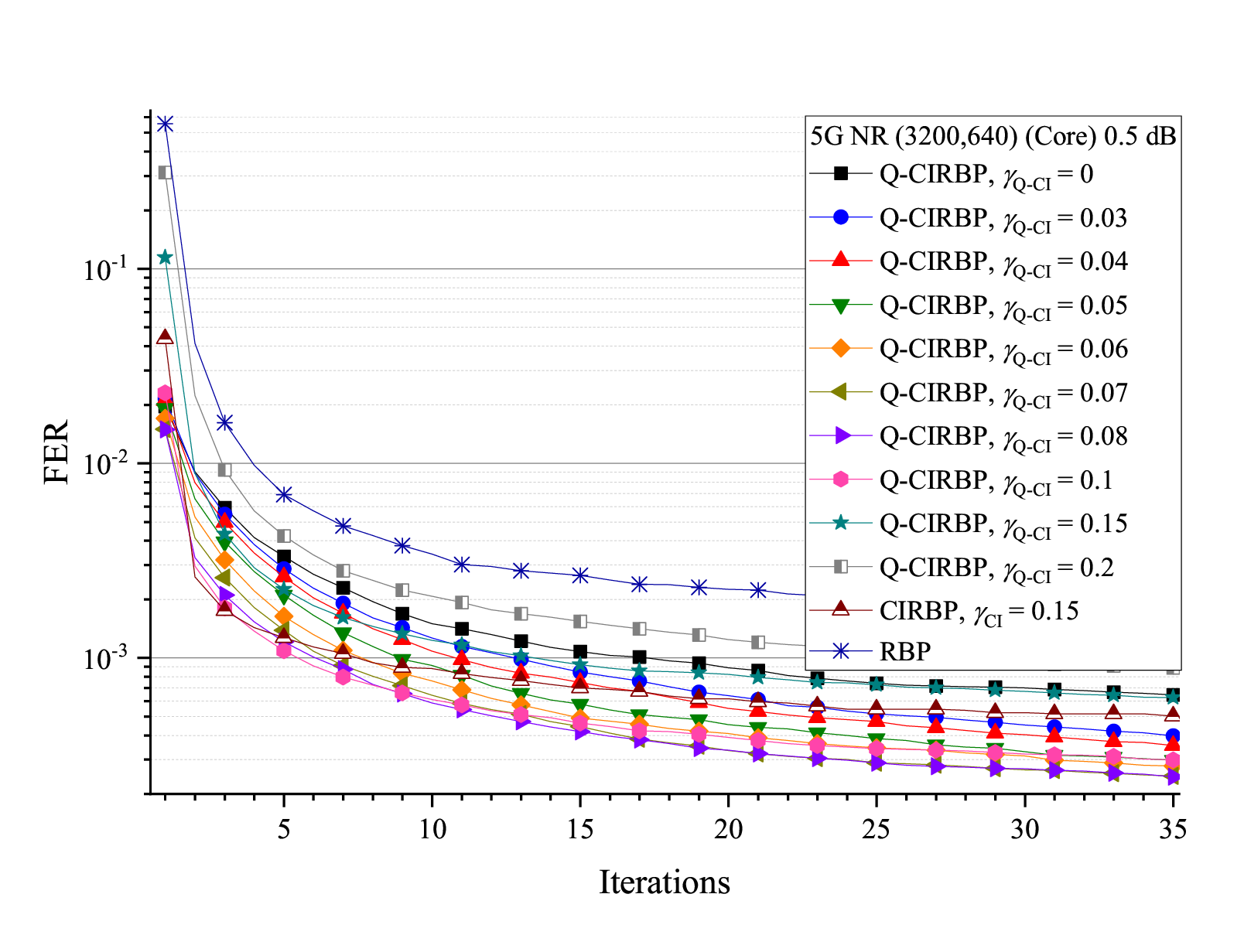}%
    \label{ITEvsFER_NR3200_EWIDS_QCI_gamma}}
    \caption{FER convergence behavior of the proposed Q-CIRBP decoder
    for different $\gamma_{\mathrm{Q\text{-}CI}}$ values
    with the rate-$1/2$ M-1008, rate-$1/2$ W-1944, and rate-$1/5$ N-3200 codes.}
    \label{Fig:QCI_gamma}
\end{figure}

Figures~\ref{Fig:ECI_gamma} and \ref{Fig:QCI_gamma}
show the FER convergence behavior of the
proposed E-CIRBP and Q-CIRBP decoders
for different threshold values
$\gamma_{\mathrm{E\text{-}CI}}$ and $\gamma_{\mathrm{Q\text{-}CI}}$.
The values of $E_b/N_0$ are set to $2.75$ dB, $1.75$ dB, and $0.5$ dB
for the M-1008, W-1944, and N-3200 codes, respectively.
The performance of the RBP decoder and that of the CIRBP decoder
with its optimal threshold are also included for reference.

Figure~\ref{ITEvsFER_MK1008_EWIDS_ECI_gamma} shows that,
for $0\leq\gamma_{\mathrm{E\text{-}CI}}\leq0.3$,
different choices of $\gamma_{\mathrm{E\text{-}CI}}$ lead to different tradeoffs
between the FER performance in the early and later decoding iterations
for the M-1008 code.
Among the tested values,
$\gamma_{\mathrm{E\text{-}CI}}=0.15$ generally provides the best performance
for almost all the observed iterations.
A similar tradeoff is observed for the W-1944 code,
although the favorable balance shifts to $\gamma_{\mathrm{E\text{-}CI}}=0.1$,
as seen in Figure~\ref{ITEvsFER_WIFI1944_EWIDS_ECI_gamma}.
For the Q-CIRBP decoder,
the results in Figure~\ref{Fig:QCI_gamma} suggest that suitable values of
$\gamma_{\mathrm{Q\text{-}CI}}$ lie approximately between $0$ and $0.1$
for both the M-1008 and W-1944 codes.
Among the tested values,
$\gamma_{\mathrm{Q\text{-}CI}}=0.03$ provides the lowest FER for $I>1$
in both cases and is therefore selected for the subsequent comparisons.
For the N-3200 code,
the results in
Figure~\ref{ITEvsFER_NR3200_EWIDS_ECI_gamma} indicate that
different threshold values lead to different convergence behavior
of the E-CIRBP decoder.
Among the tested values,
$\gamma_{\mathrm{E\text{-}CI}}=0.4$
generally provides the best overall FER performance
over the observed decoding iterations
and is therefore adopted in the subsequent comparisons.
For the Q-CIRBP decoder,
Figure~\ref{ITEvsFER_NR3200_EWIDS_QCI_gamma}
shows that
$\gamma_{\mathrm{Q\text{-}CI}}=0.08$
generally yields the best FER performance
among the threshold values tested
and is selected as the optimal threshold.

Based on these observations, unless otherwise specified,
$\gamma_{\mathrm{E\text{-}CI}}=0.15$, $0.1$, and $0.4$
are used for the E-CIRBP decoder
with the M-1008, W-1944, and N-3200 codes, respectively,
whereas $\gamma_{\mathrm{Q\text{-}CI}}=0.03$, $0.03$, and $0.08$
are respectively chosen for the Q-CIRBP decoder with the three codes.

\begin{figure}[!htbp]
    \centering
    \subfloat[FER performance versus $E_b/N_0$; $I_{\max}=3$.]
    {\includegraphics[width=3.5in]{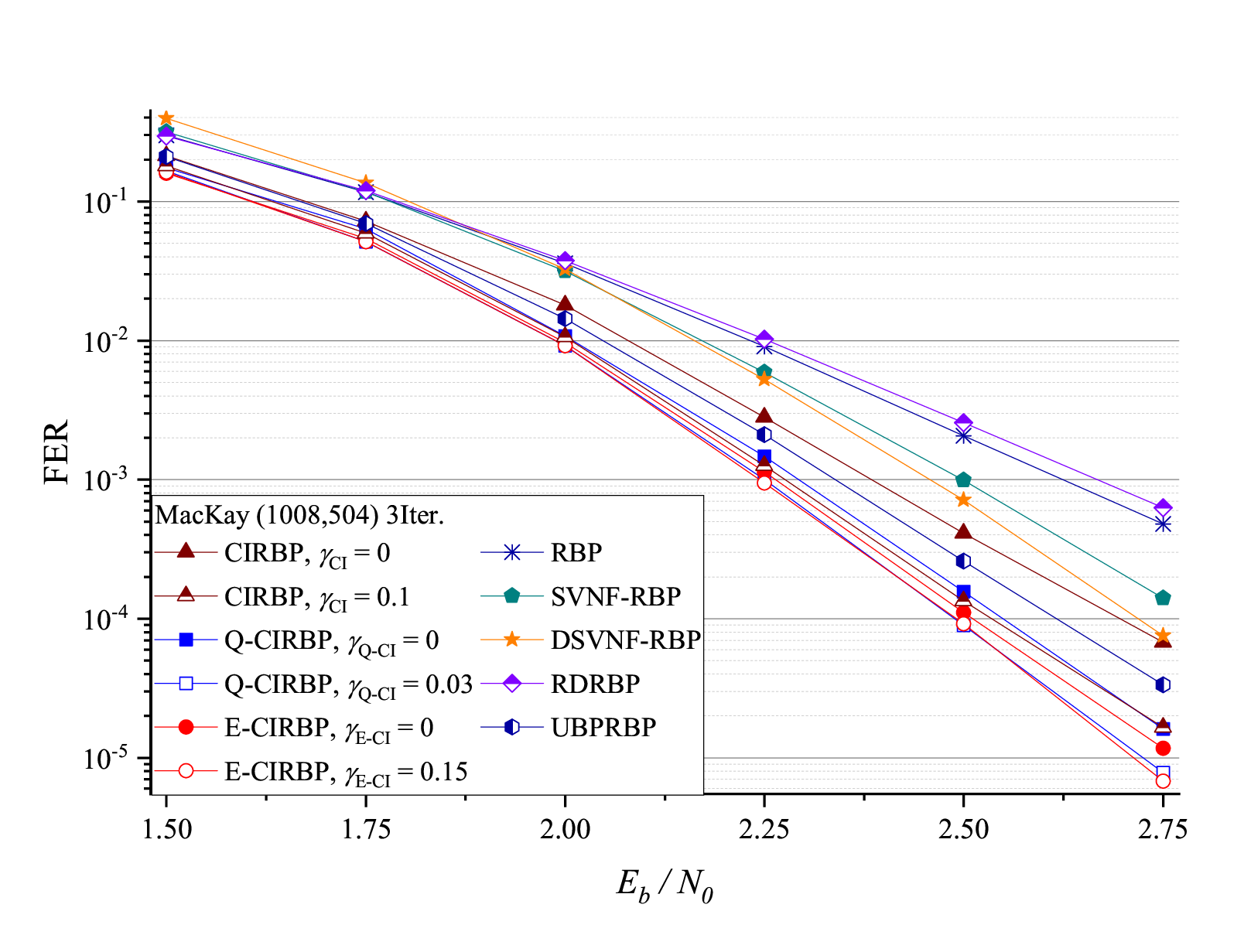}%
    \label{SNRvsFER_MK1008_EWIDS_3iter}}
    \hspace{0.00\textwidth}
    \subfloat[FER convergence behavior; $E_b/N_0=2.75$ dB.]
    {\includegraphics[width=3.5in]{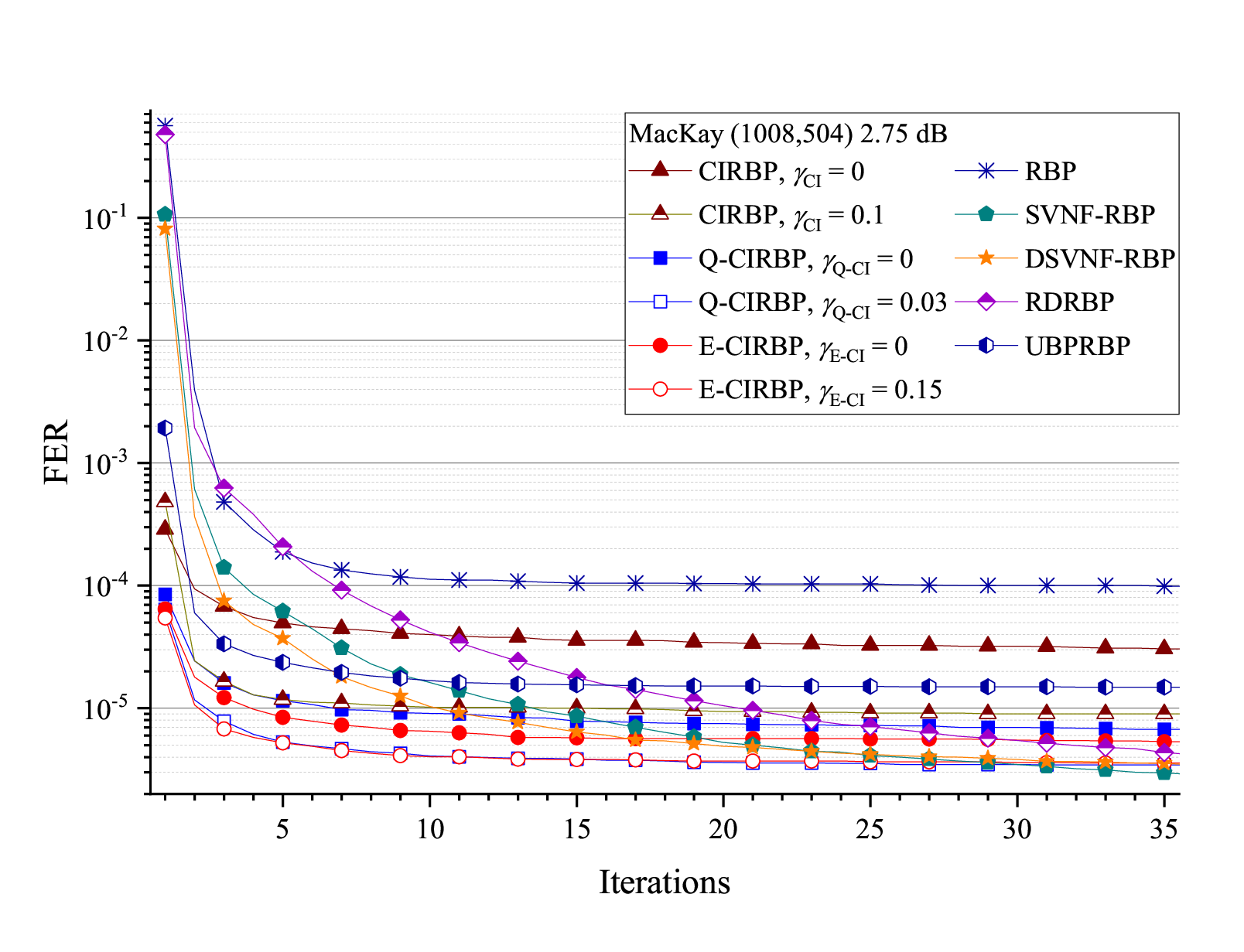}%
    \label{ITEvsFER_MK1008_EWIDS_2.75dB}}
    \caption{FER performance and convergence behavior of various IDS-based BP decoders for the rate-$1/2$ M-1008 code.}
    \label{Fig:EW_M1008}
\end{figure}

Figure~\ref{SNRvsFER_MK1008_EWIDS_3iter} compares the FER performance
of the proposed E-CIRBP and Q-CIRBP decoders
with the original CIRBP decoder and the other IDS-based BP decoders
for the M-1008 code at $I_{\max}=3$.
It can be seen that the E-CIRBP decoder achieves the lowest FER
over the entire simulated $E_b/N_0$ range.
At FER $\approx2\times10^{-5}$,
the E-CIRBP decoder provides a gain of approximately $0.1$ dB
over the CIRBP algorithm with the optimal threshold $\gamma_{\mathrm{CI}}=0.1$,
as determined through simulation.
At FER $\approx10^{-3}$,
it also provides a gain of at least $0.09$ dB
over the other reference decoders.
The Q-CIRBP decoder provides similar improvements.
At FER $\approx2\times10^{-5}$,
it achieves a gain of approximately $0.07$ dB
over the CIRBP decoder with $\gamma_{\mathrm{CI}}=0.1$,
and at FER $\approx10^{-3}$,
it provides a gain of approximately $0.09$ dB over the UBPRBP algorithm.

Figure~\ref{ITEvsFER_MK1008_EWIDS_2.75dB} further presents
the FER convergence behavior at $E_b/N_0=2.75$ dB
over $1\leq I\leq35$.
The results indicate that the E-CIRBP decoder gives the lowest FER
among the compared decoders for approximately the first $30$ iterations.
The Q-CIRBP decoder follows a similar trend,
and its FER remains close to that of the E-CIRBP algorithm
throughout the observed iteration range.
For the later iterations,
most of the decoders are already close to convergence,
whereas the FER of the SVNF-RBP algorithm continues to decrease
and becomes competitive near the end of the observed range.
Nevertheless, the advantage of the proposed E-CIRBP and Q-CIRBP decoders
is particularly evident during the first few decoding iterations.
Specifically, when the first iteration is completed,
the FERs achieved by E-CIRBP and Q-CIRBP
are approximately $11\%$ and $13\%$, respectively,
of that achieved by CIRBP with its optimal threshold $\gamma_{\mathrm{CI}}=0.1$.

\begin{figure}[!htbp]
    \centering
    \subfloat[FER performance versus $E_b/N_0$; $I_{\max}=3$.]
    {\includegraphics[width=3.5in]{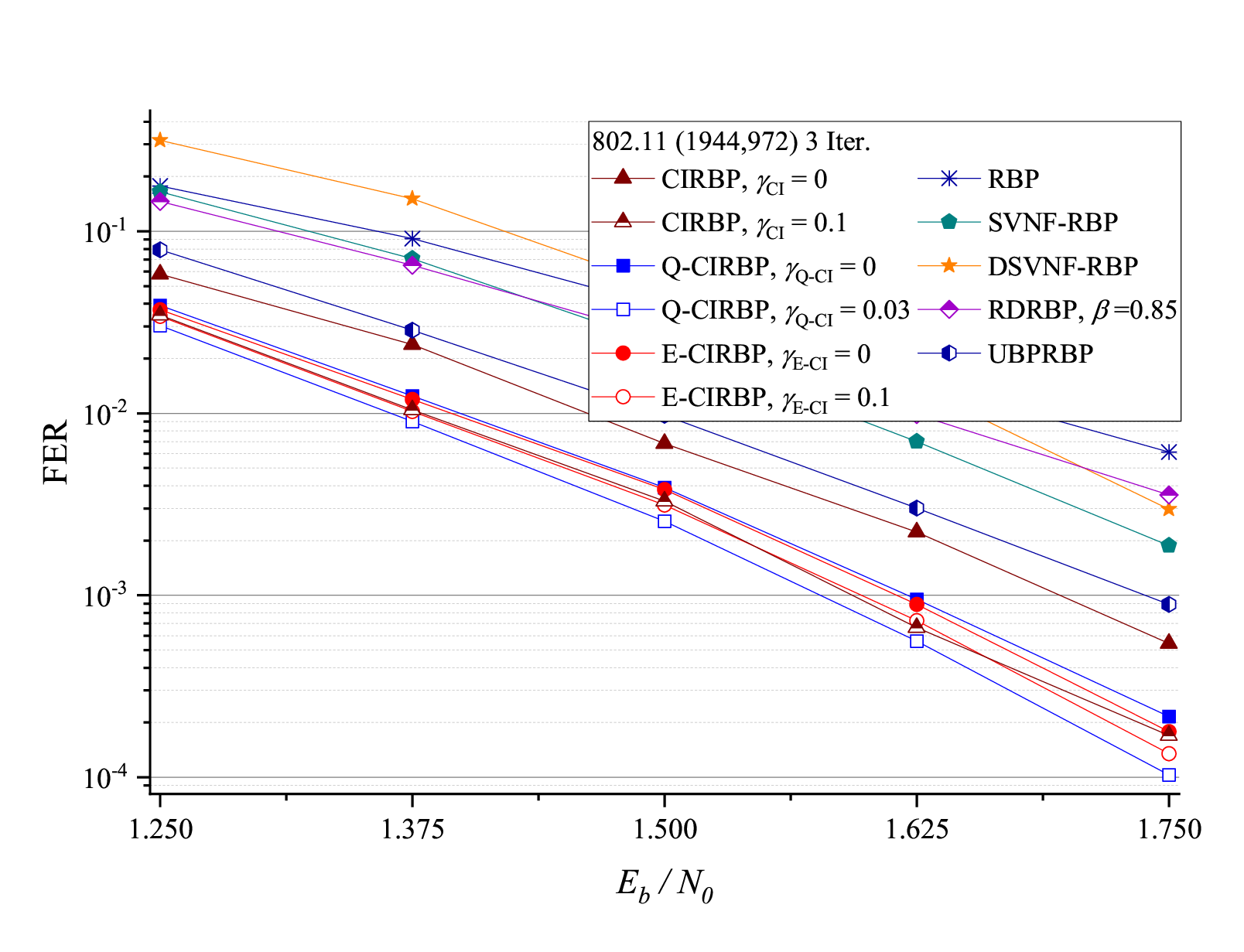}%
    \label{SNRvsFER_WIFI1944_EWIDS_3iter}}
    \hspace{0.00\textwidth}
    \subfloat[FER convergence behavior; $E_b/N_0=1.75$ dB.]
    {\includegraphics[width=3.5in]{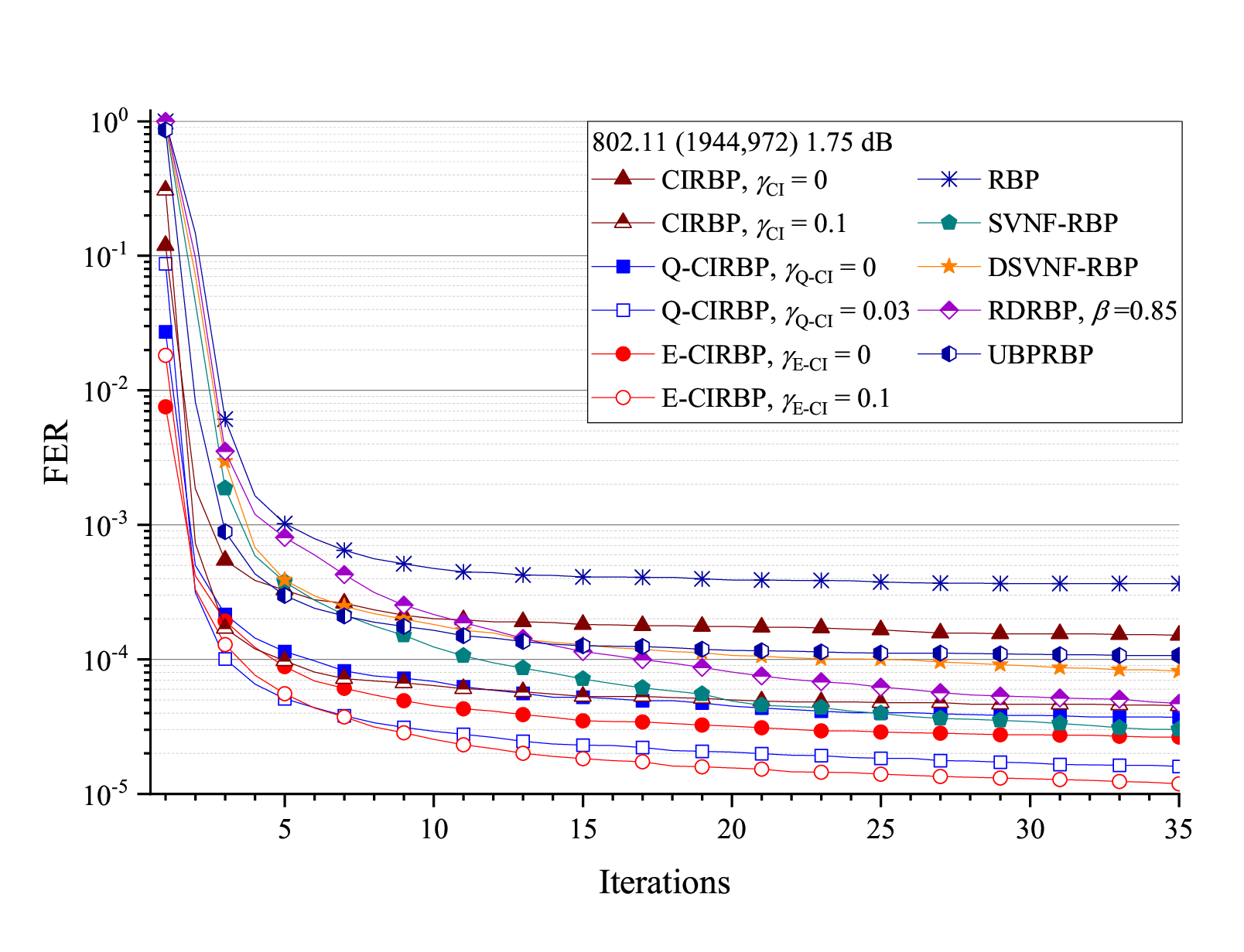}%
    \label{ITEvsFER_WIFI1944_EWIDS_1.75dB}}
    \caption{FER performance and convergence behavior of various IDS-based BP decoders for the rate-$1/2$ W-1944 code.}
    \label{Fig:EW_W1944}
\end{figure}

For the W-1944 code,
Figure~\ref{SNRvsFER_WIFI1944_EWIDS_3iter}
compares the FER performance of the considered decoders
at $I_{\max}=3$.
With its optimal threshold $\gamma_{\mathrm{CI}}=0.1$,
the CIRBP decoder already outperforms the other reference IDS-based BP decoders.
For example, it provides a gain of approximately $0.15$ dB
over UBPRBP at FER $\approx10^{-3}$,
while the proposed E-CIRBP and Q-CIRBP decoders
provide further FER improvements over CIRBP
throughout the simulated $E_b/N_0$ range.
The corresponding convergence behavior is shown in
Figure~\ref{ITEvsFER_WIFI1944_EWIDS_1.75dB}
for $E_b/N_0=1.75$ dB and $1\leq I\leq35$.
It can be observed that both the E-CIRBP and Q-CIRBP algorithms
achieve lower FER than the CIRBP decoder with its optimal threshold $\gamma_{\mathrm{CI}}=0.1$
and the other reference IDS-based BP decoders
throughout the observed iteration range.
This advantage is particularly evident
at the beginning of the decoding process.
Specifically, when the first iteration is completed,
the FER achieved by E-CIRBP is approximately one-tenth
of that achieved by CIRBP,
whereas that of Q-CIRBP is approximately one-third.

\begin{figure}[!htbp]
    \centering
    \subfloat[FER performance versus $E_b/N_0$; $I_{\max}=20$.]
    {\includegraphics[width=3.5in]{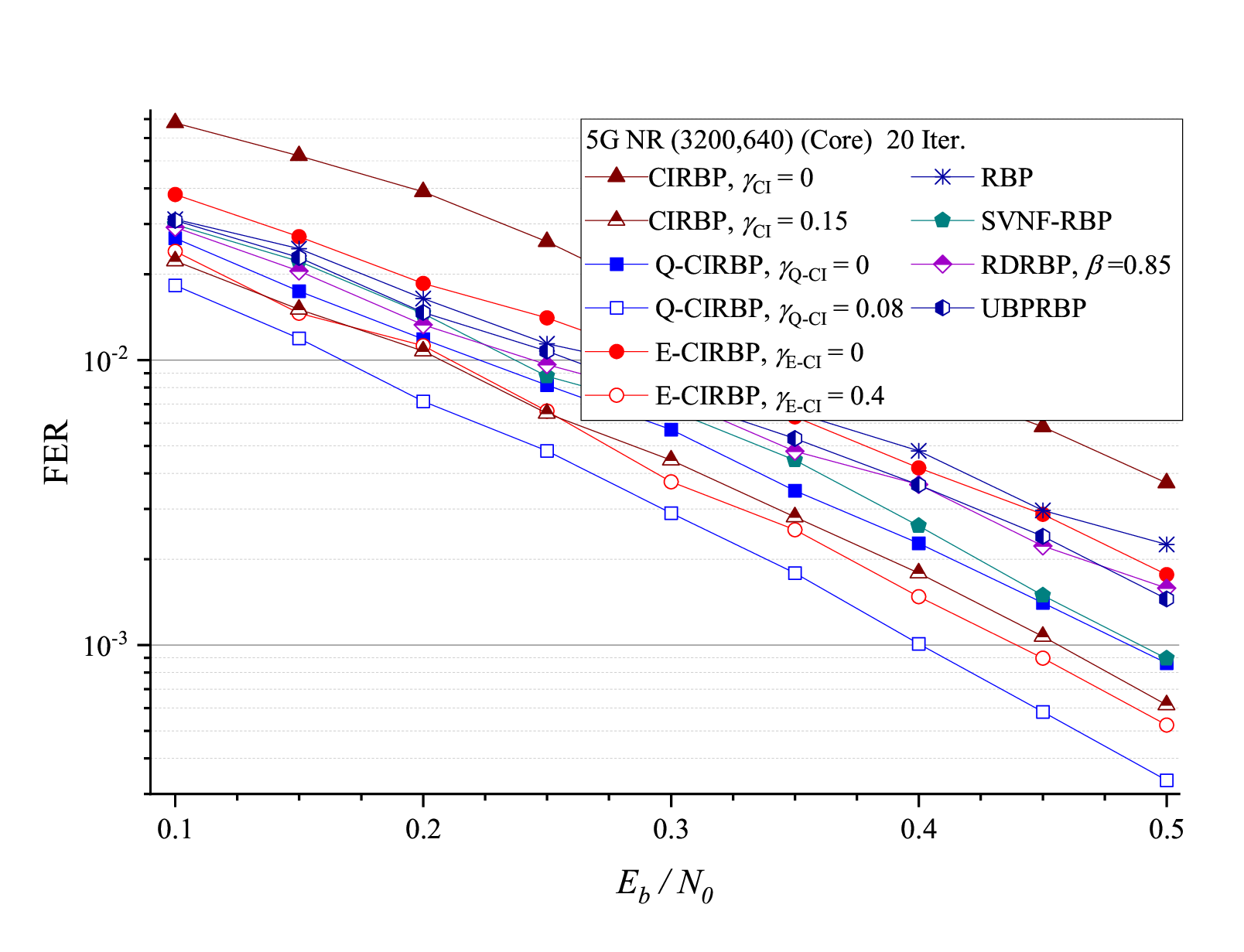}%
    \label{SNRvsFER_NR3200_EWIDS_20iter}}
    \hspace{0.00\textwidth}
    \subfloat[FER convergence behavior; $E_b/N_0=0.5$ dB.]
    {\includegraphics[width=3.5in]{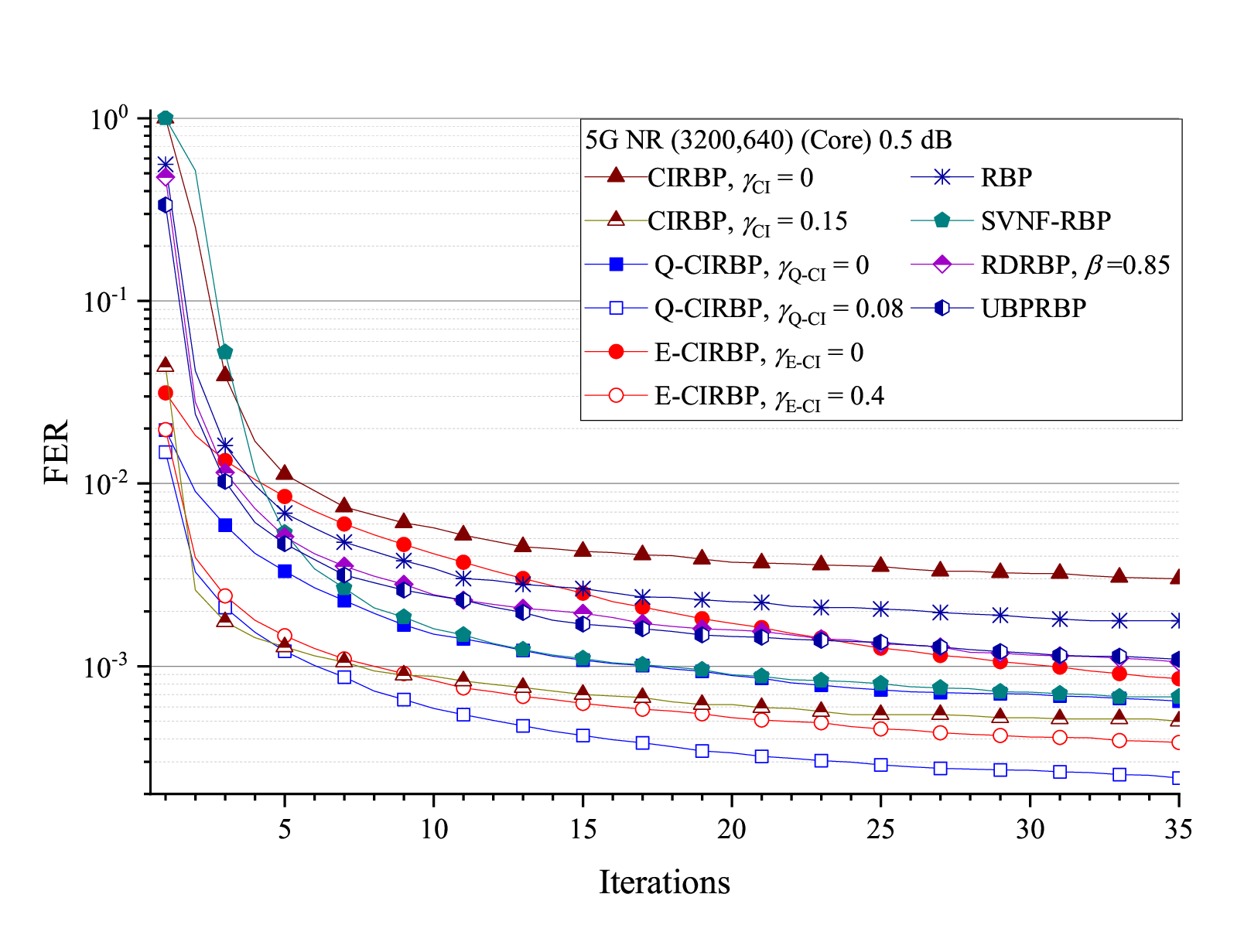}%
    \label{ITEvsFER_NR3200_EWIDS_0.5dB}}
    \caption{FER performance and convergence behavior of various IDS-based BP decoders for the rate-$1/5$ N-3200 code.}
    \label{Fig:EW_N3200}
\end{figure}

For the N-3200 code,
Figure~\ref{SNRvsFER_NR3200_EWIDS_20iter}
compares the FER performance of the considered decoders
at $I_{\max}=20$.
The Q-CIRBP decoder provides superior performance
over the entire simulated $E_b/N_0$ range.
The E-CIRBP decoder also outperforms
the CIRBP decoder with the optimized threshold
and the compared IDS-based BP decoders
throughout the simulated range.
Figure~\ref{ITEvsFER_NR3200_EWIDS_0.5dB}
further presents the FER convergence behavior
at $E_b/N_0=0.5$ dB over $1\leq I\leq35$.
The Q-CIRBP decoder yields the lowest FER
at the first decoding iteration
and for most of the subsequent decoding iterations.
The E-CIRBP decoder also achieves better FER performance
compared with the CIRBP decoder
in the later decoding iterations.

Figures~\ref{ITEvsFER_MK1008_EWIDS_2.75dB} and
\ref{ITEvsFER_WIFI1944_EWIDS_1.75dB}
also reveal an interesting tradeoff between the initial and
converged FER performance of the original CIRBP decoder
under different threshold settings.
When $\gamma_{\mathrm{CI}}=0$ is used,
the global maximum-residual search is never activated during decoding.
This setting results in a considerably lower FER
at the first decoding iteration,
but the converged FER is less favorable than that obtained with the
optimized nonzero threshold.
In contrast, CIRBP with
the optimized nonzero threshold
shows less favorable FER in the beginning,
but provides better performance in the later iterations.
This observation indicates that the selection of
$\gamma_{\mathrm{CI}}$
may involve a tradeoff among several requirements,
such as decoding latency, converged FER performance,
and computational complexity.
The proposed E-CIRBP and Q-CIRBP decoders, however,
exhibit less sensitivity to this tradeoff.
In particular, when
$\gamma_{\mathrm{E\text{-}CI}}$ and
$\gamma_{\mathrm{Q\text{-}CI}}$
are both set to zero,
the two proposed decoders still provide favorable FER
at the first decoding iteration,
while their converged performance remains satisfactory.
More importantly,
Figures~\ref{ITEvsFER_MK1008_EWIDS_2.75dB} and
\ref{ITEvsFER_WIFI1944_EWIDS_1.75dB}
show that, even with zero thresholds,
the proposed E-CIRBP and Q-CIRBP decoders can still outperform
the CIRBP decoder using its
optimized nonzero threshold
throughout the observed iteration range.
These results suggest that,
compared with the original CI metric,
the proposed E-CI and Q-CI metrics may provide
improved FER performance with reduced sensitivity
to threshold selection.
However, for the N-3200 code,
the selected nonzero thresholds are more important
for achieving improved FER performance
with the proposed metrics.
As shown in Figure~\ref{ITEvsFER_NR3200_EWIDS_0.5dB},
using these thresholds substantially improves
the FER performance of both the E-CIRBP and Q-CIRBP decoders
relative to their corresponding zero-threshold settings
and enables them to achieve favorable performance
relative to the CIRBP decoder with its optimized nonzero threshold.
This observation indicates that the sensitivity
to threshold selection can be code-dependent.

\begin{figure}[!htbp]
    \centering
    \subfloat[FER performance versus $E_b/N_0$; $I_{\max}=10$.]
    {\includegraphics[width=3.5in]{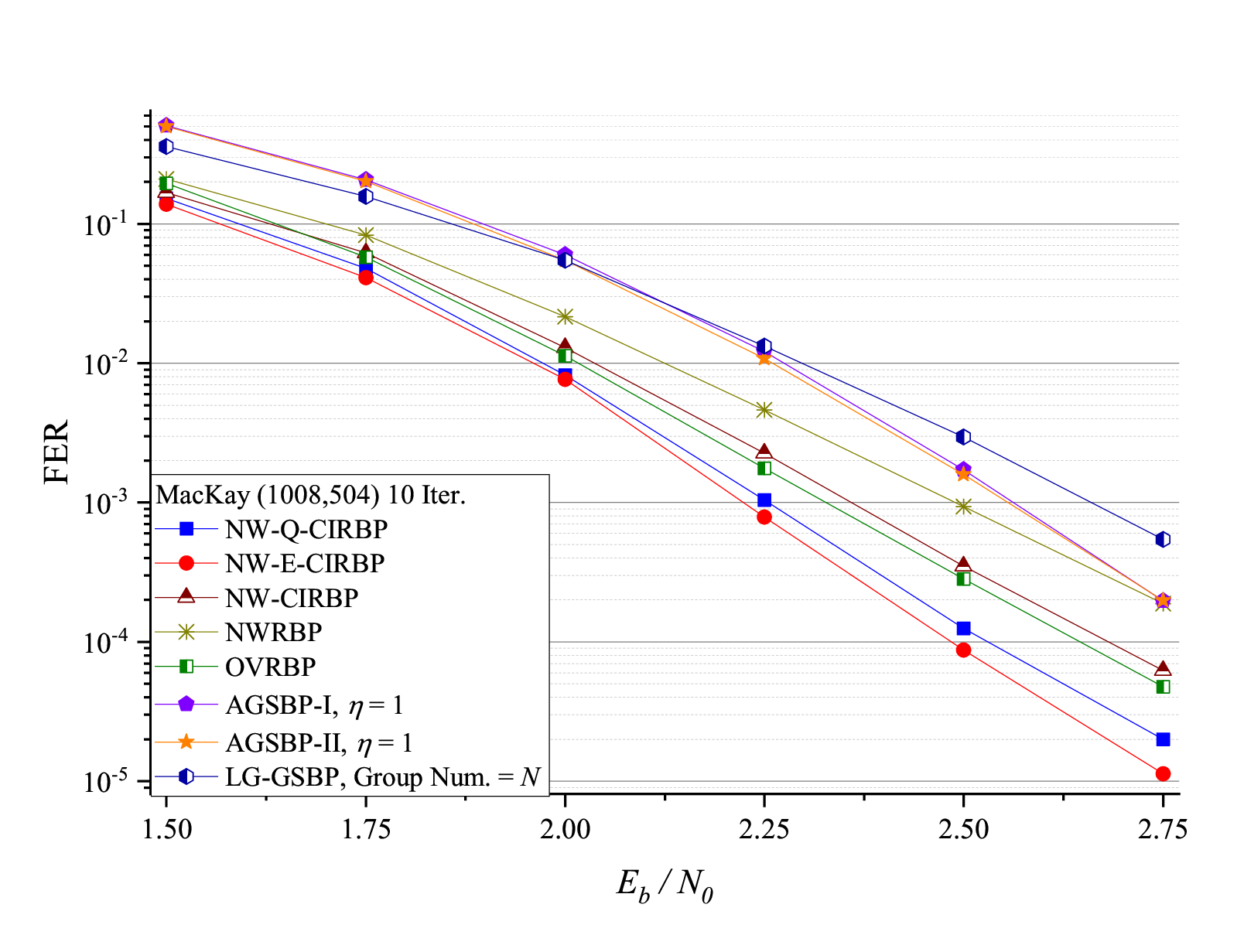}%
    \label{SNRvsFER_MK1008_NWIDS_10iter}}
    \hspace{0.00\textwidth}
    \subfloat[FER convergence behavior; $E_b/N_0=2.75$ dB.]
    {\includegraphics[width=3.5in]{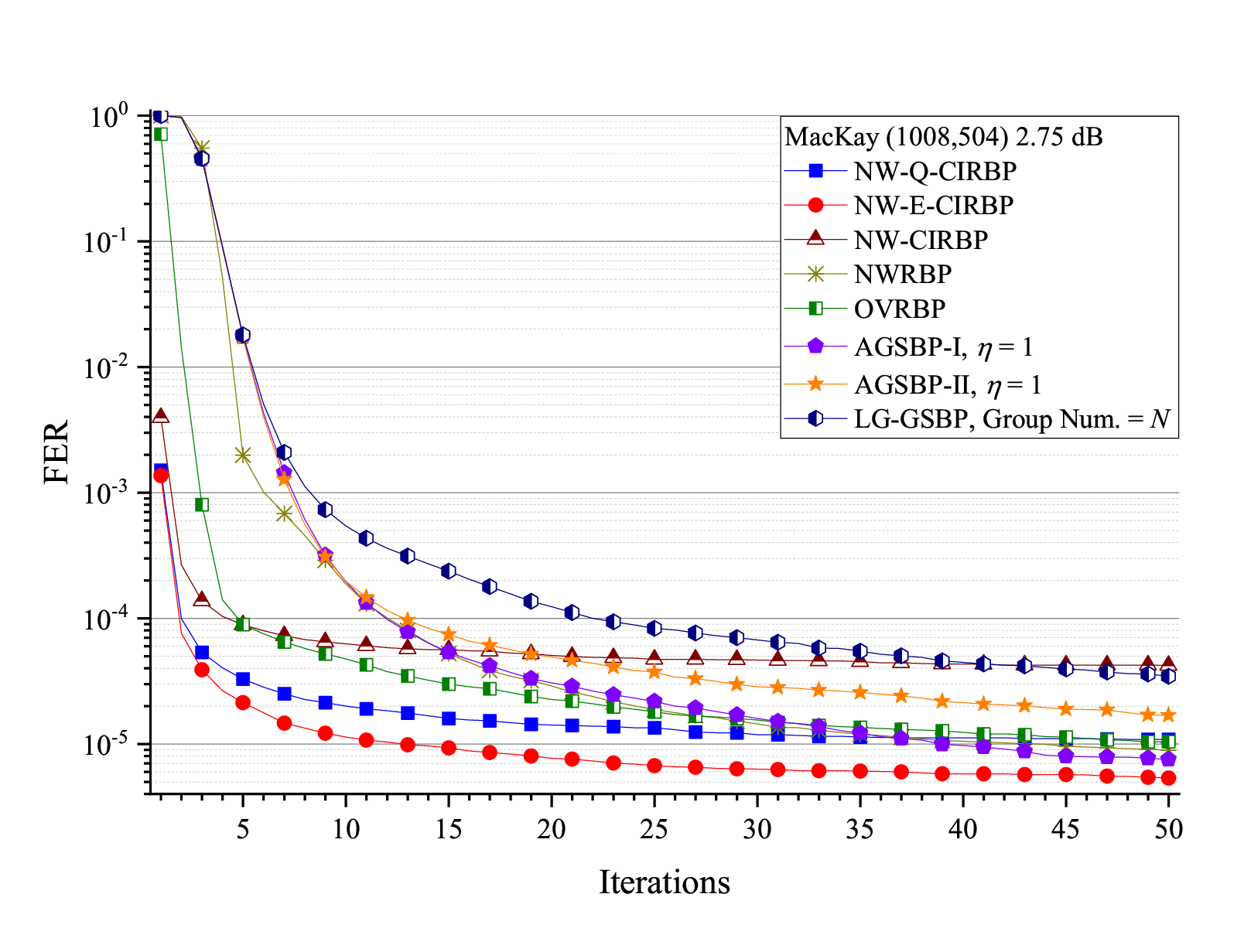}%
    \label{ITEvsFER_MK1008_NWIDS_2.75dB}}
    \caption{FER performance and convergence behavior of various node-wise BP decoders for the rate-$1/2$ M-1008 code.}
    \label{Fig:NW_M1008}
\end{figure}

\begin{figure}[!htbp]
    \centering
    \subfloat[FER performance versus $E_b/N_0$; $I_{\max}=10$.]
    {\includegraphics[width=3.5in]{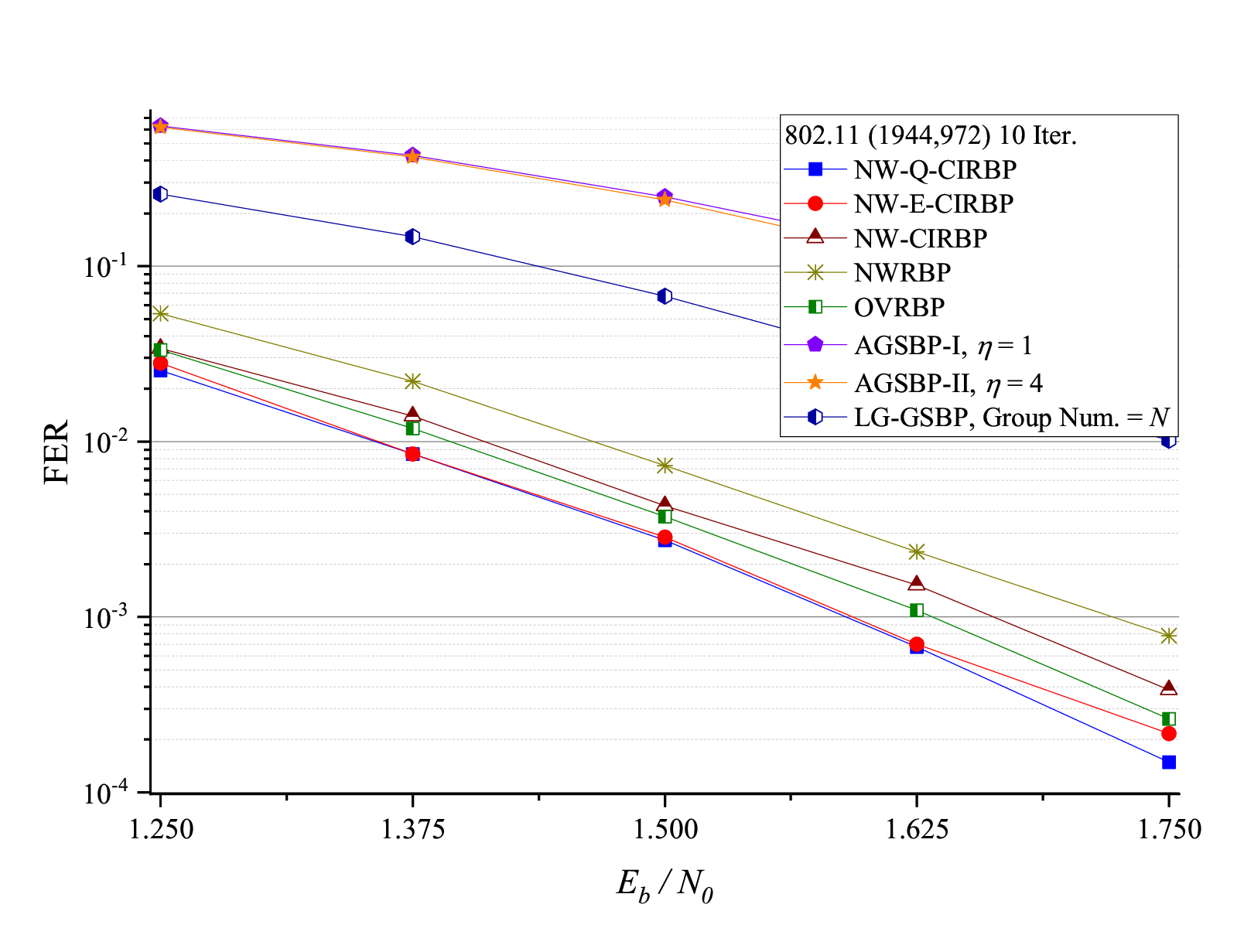}%
    \label{SNRvsFER_WIFI1944_NWIDS_10iter}}
    \hspace{0.00\textwidth}
    \subfloat[FER convergence behavior; $E_b/N_0=1.75$ dB.]
    {\includegraphics[width=3.5in]{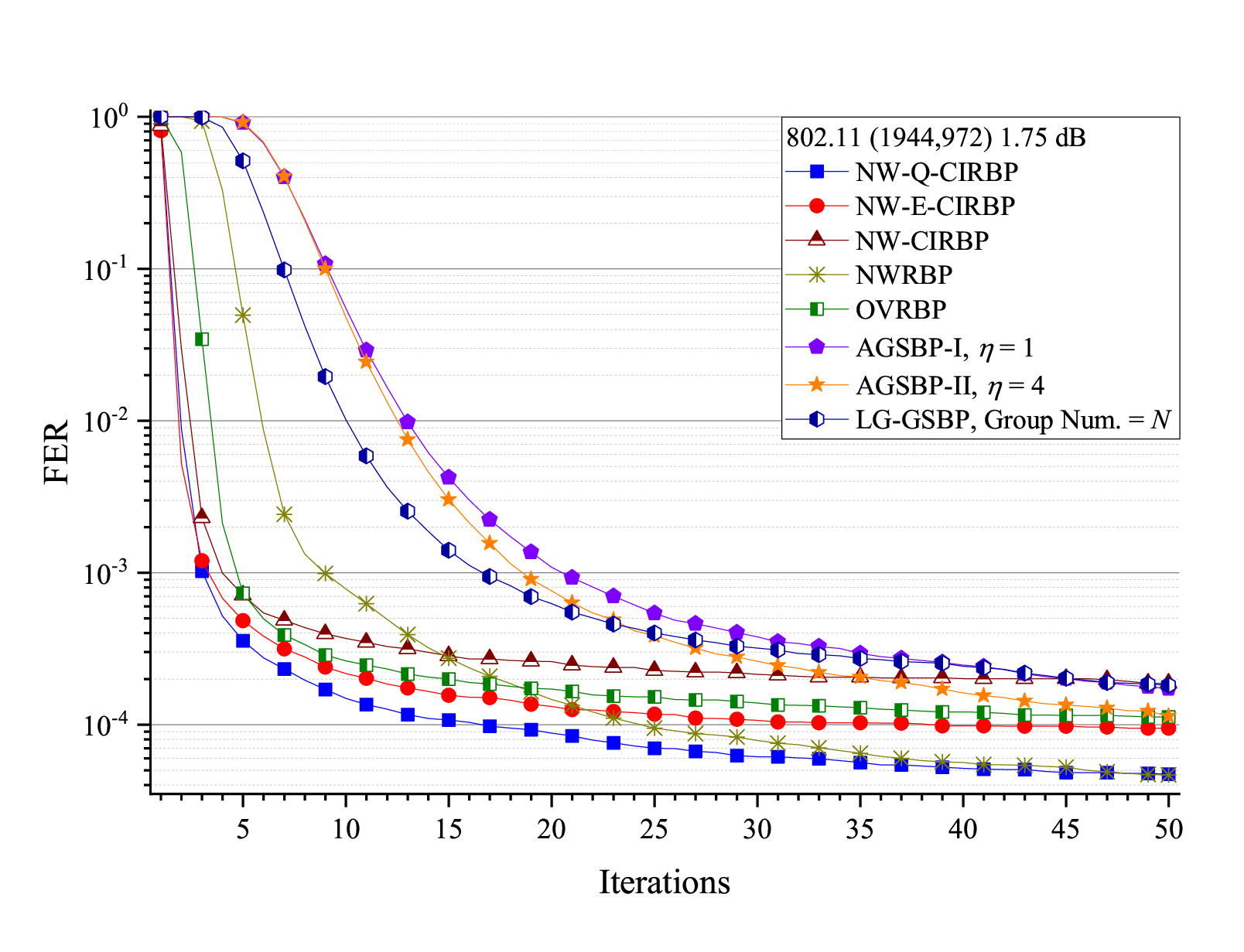}%
    \label{ITEvsFER_WIFI1944_NWIDS_1.75dB}}
    \caption{FER performance and convergence behavior of various node-wise BP decoders for the rate-$1/2$ W-1944 code.}
    \label{Fig:NW_W1944}
\end{figure}

\begin{figure}[!htbp]
    \centering
    \subfloat[FER performance versus $E_b/N_0$; $I_{\max}=10$.]
    {\includegraphics[width=3.5in]{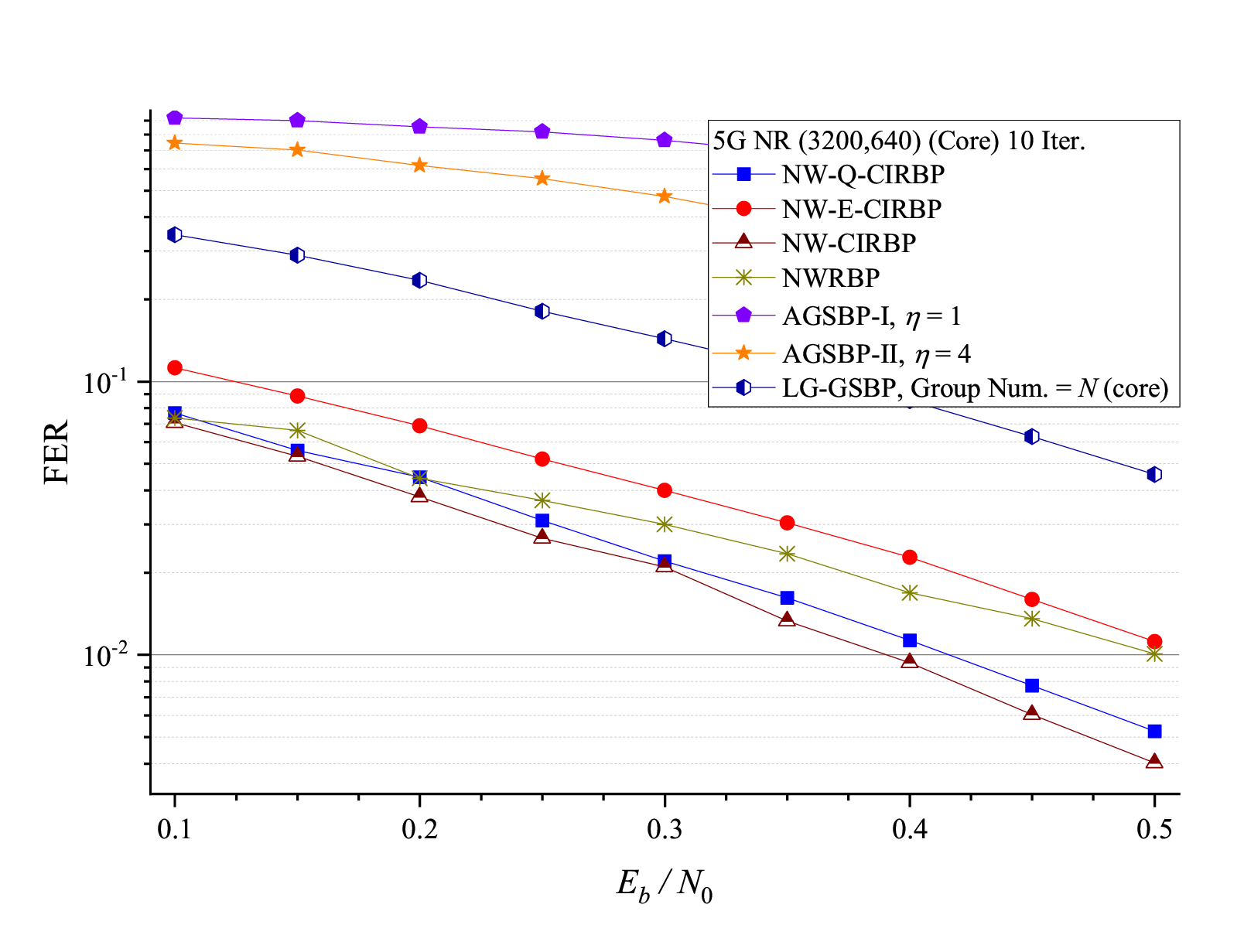}%
    \label{SNRvsFER_NR3200_NWIDS_10iter}}
     \hspace{0.00\textwidth}
    \subfloat[FER convergence behavior; $E_b/N_0=0.5$ dB.]
    {\includegraphics[width=3.5in]{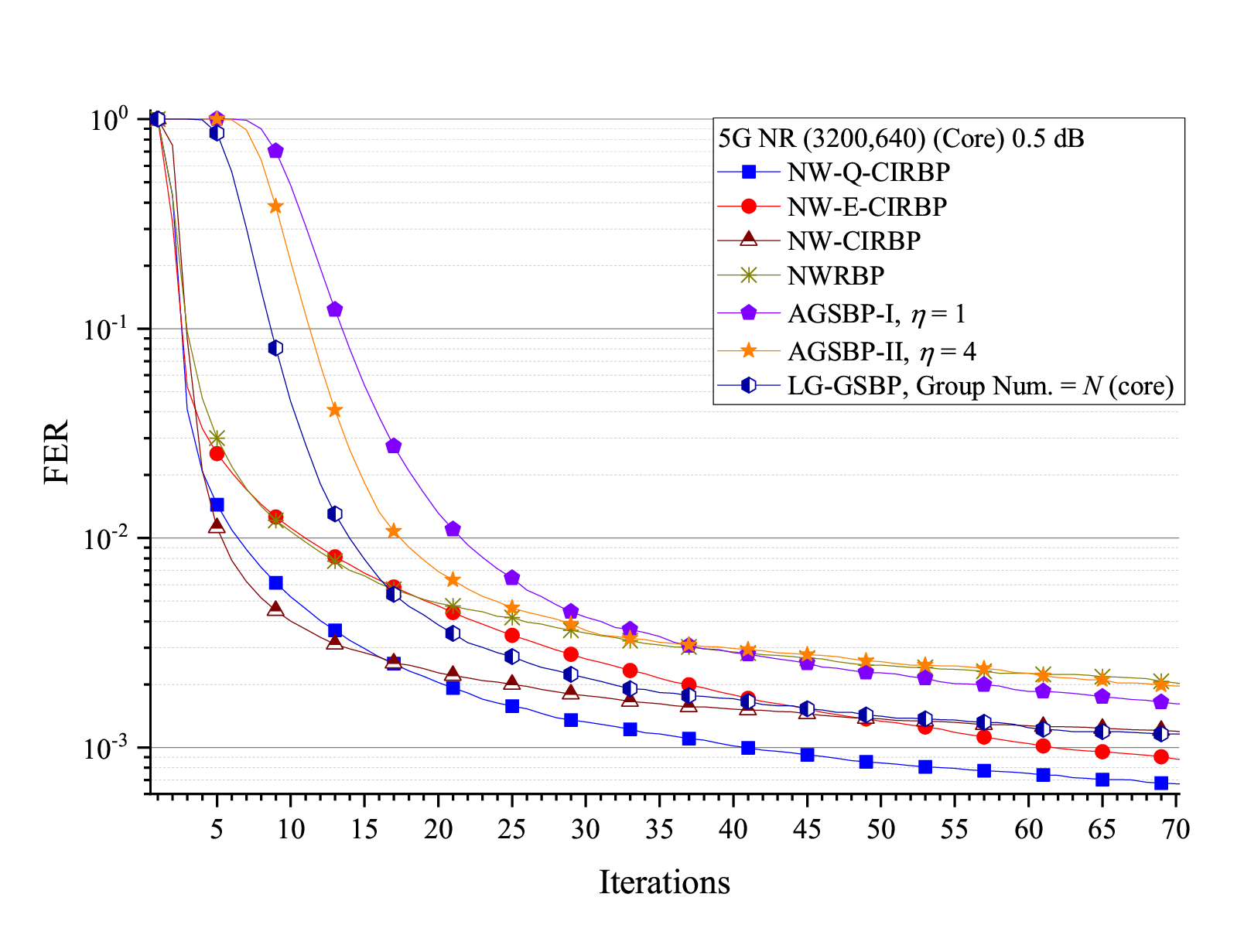}%
    \label{ITEvsFER_NR3200_NWIDS_0.5dB}}
    \caption{FER performance and convergence behavior of various node-wise BP decoders for the rate-$1/5$ N-3200 code.}
    \label{Fig:NW_N3200}
\end{figure}

Figures~\ref{SNRvsFER_MK1008_NWIDS_10iter},
\ref{SNRvsFER_WIFI1944_NWIDS_10iter}, and
\ref{SNRvsFER_NR3200_NWIDS_10iter}
present the FER performance of various BP decoders
with multiple-edge updates
at $I_{\max}=10$ for the M-1008, W-1944, and N-3200 codes.
The NW-CIRBP decoder employs the original CI metric,
whereas the NW-E-CIRBP and NW-Q-CIRBP decoders
employ the proposed E-CI and Q-CI metrics.
Figure~\ref{SNRvsFER_MK1008_NWIDS_10iter}
demonstrates that the NW-E-CIRBP decoder
outperforms all the other compared decoders for the M-1008 code.
At FER $\approx5\times10^{-5}$,
it provides gains of approximately $0.18$ dB over the OVRBP decoder
and $0.20$ dB over the NW-CIRBP decoder.
The NW-Q-CIRBP decoder also outperforms these two decoders
at the same FER,
providing gains of approximately $0.13$ dB over the OVRBP decoder
and $0.17$ dB over the NW-CIRBP decoder.
For the W-1944 code,
the results in Figure~\ref{SNRvsFER_WIFI1944_NWIDS_10iter}
indicate that the NW-E-CIRBP decoder achieves the lowest FER
when $E_b/N_0\leq1.625$ dB,
whereas the NW-Q-CIRBP decoder provides the best performance
at $E_b/N_0=1.75$ dB.
Nevertheless,
both the proposed NW-E-CIRBP and NW-Q-CIRBP algorithms
outperform the other compared decoders
over the entire simulated $E_b/N_0$ range.
For the N-3200 code,
Figure~\ref{SNRvsFER_NR3200_NWIDS_10iter}
shows a different relative performance
among the three CI-based node-wise decoders.
At $I_{\max}=10$,
the NW-CIRBP decoder provides the best FER performance
over the simulated $E_b/N_0$ range,
while the NW-Q-CIRBP decoder provides comparable performance.
In particular, at $E_b/N_0=0.5$ dB,
the FER of the NW-Q-CIRBP decoder
is approximately $50\%$ lower than that of the NWRBP decoder.

The FER convergence behavior of
the considered BP decoders with multiple-edge updates
for the M-1008 code is presented in
Figure~\ref{ITEvsFER_MK1008_NWIDS_2.75dB},
where $E_b/N_0=2.75$ dB.
It can be observed that, when $I<5$,
the NW-CIRBP, NW-E-CIRBP, and NW-Q-CIRBP decoders
all provide substantial FER improvements
over the NWRBP and OVRBP decoders.
Furthermore, for $I>1$,
the proposed NW-E-CIRBP and NW-Q-CIRBP decoders
significantly outperform the NW-CIRBP decoder,
and this advantage is maintained until convergence.
Among the compared decoders,
the proposed NW-E-CIRBP decoder achieves the lowest FER.
By $I=50$,
its FER is reduced to approximately $1/10$
of that achieved by the NW-CIRBP decoder.
A similar advantage of the proposed metrics
can also be observed for the W-1944 code
in Figure~\ref{ITEvsFER_WIFI1944_NWIDS_1.75dB},
where $E_b/N_0=1.75$ dB.
During the first $10$ iterations,
both the NW-E-CIRBP and NW-Q-CIRBP decoders
provide noticeably lower FER
than that achieved by the NW-CIRBP, NWRBP, and OVRBP decoders.
As the decoding proceeds,
the performance differences gradually decrease.
For $I>35$,
the NW-E-CIRBP and OVRBP decoders achieve similar FER performance,
while the NW-Q-CIRBP decoder performs comparably
to the NWRBP decoder.
For the N-3200 code,
the convergence behavior at $E_b/N_0=0.5$ dB
is shown in Figure~\ref{ITEvsFER_NR3200_NWIDS_0.5dB}
over $1\leq I\leq70$.
The NW-CIRBP decoder provides better FER performance
during the earlier decoding iterations.
As the decoding proceeds,
the FER of the NW-Q-CIRBP decoder continues to decrease
and becomes lower than that of the NW-CIRBP decoder
after approximately $20$ iterations.
The FER of the NW-E-CIRBP decoder also continues to decrease
and becomes lower than that of the NW-CIRBP decoder
in the later observed iterations.
By $I=70$,
both the NW-E-CIRBP and NW-Q-CIRBP decoders
provide lower FER than the NW-CIRBP decoder.
Across the three codes,
the LG-GSBP and AGSBP decoders
generally exhibit slower FER reduction
during the early decoding iterations,
while the performance differences become smaller
as decoding proceeds.

Taken together,
these performance results suggest that
the proposed E-CI and Q-CI metrics
may be more effective than the original CI metric
for determining the update priorities of VNs
under both the procedure in which a single C2V message
is propagated at each update,
as described in Algorithm~\ref{alg:CIRBP_general},
and the node-wise procedure in which multiple C2V messages
are propagated at each update,
as described in Algorithm~\ref{alg:NW_CIRBP_general}
,
with their relative FER advantages depending on
the code, updating strategy, and number of decoding iterations.

\subsection{Complexity Considerations}
\label{Sec4_B}

Table~\ref{tab:per_update_complexity}
summarizes the computational requirements per scheduling update
for the proposed decoders and the reference IDS-based BP decoders.
The comparison includes C2V message propagations,
V2C message updates,
C2V precomputations,
the comparisons required to select the VN and/or edge to be updated,
and, for the CI-based decoders,
the number of CI-related metric updates.
In the table,
$\bar d_v=E/N$ and $\bar d_c=E/M$
denote the average VN and CN degrees, respectively,
and $\widetilde M$ denotes the number of unmuted CNs
at the current update.
In addition, for the GSBP-based decoders,
message precomputation and scheduling-comparison costs
are absent (for LG-GSBP \cite{LG-GSBP})
or negligible (for AGSBP \cite{AGSBP})
relative to those of the IDS-based decoders,
and their computational requirements are therefore
treated as those of conventional SBP decoding,
involving only V2C and C2V message updates.

\begin{table}[!htbp]
\caption{Summary of the required computational complexity per update for the proposed and reference IDS-based BP decoders.}
\label{tab:per_update_complexity}
\centering
\normalsize
\renewcommand{\arraystretch}{1.4}
\setlength{\tabcolsep}{4pt}
\resizebox{\linewidth}{!}{%
\begin{tabular}{|c|c|c|c|c|c|}
\hline
 & \begin{tabular}[c]{@{}c@{}}C2V\\Propagation\end{tabular}
 & \begin{tabular}[c]{@{}c@{}}V2C\\Update\end{tabular}
 & \begin{tabular}[c]{@{}c@{}}C2V\\Precomputation\end{tabular}
 & \begin{tabular}[c]{@{}c@{}}Comparisons\end{tabular}
 & \begin{tabular}[c]{@{}c@{}}CI-Related\\Metric Update\end{tabular}
\\ \hline
RBP
 & \multirow[c]{5}{*}{$1$}
 & \multirow[c]{5}{*}{$\bar d_v-1$}
 & \multirow[c]{5}{*}{$(\bar d_v-1)(\bar d_c-1)$}
 & $E-1$
 & \multirow[c]{5}{*}{$0$}
\\ \cline{1-1}\cline{5-5}
SVNF-RBP & & & & $\bar d_v(\bar d_c-1)-1$ &
\\ \cline{1-1}\cline{5-5}
DSVNF-RBP & & & & $\leq \bar d_v(\bar d_c-1)-1$ &
\\ \cline{1-1}\cline{5-5}
RDRBP & & & & $E-1$ &
\\ \cline{1-1}\cline{5-5}
UBPRBP & & & & $\widetilde M\bar d_c-1$ &
\\ \hline
CIRBP
 & \multirow[c]{3}{*}{$1$}
 & \multirow[c]{3}{*}{$\bar d_v-1$}
 & \multirow[c]{3}{*}{$(\bar d_v-1)(\bar d_c-1)$}
 & \multirow[c]{3}{*}{$\geq N+(\bar d_v-1)$}
 & \multirow[c]{3}{*}{$1+(\bar d_v-1)(\bar d_c-1)$}
\\ \cline{1-1}
E-CIRBP & & & & &
\\ \cline{1-1}
Q-CIRBP & & & & &
\\ \hline
NWRBP & $\bar d_c$ & $\bar d_c(\bar d_v-1)$
 & $\bar d_c(\bar d_v-1)(\bar d_c-1)$ & $E-1$ & $0$
\\ \hline
OVRBP & $\bar d_v$ & $\bar d_v$ & $\bar d_v(\bar d_c-1)$
 & $\leq 2(N-1)$ & $0$
\\ \hline
NW-CIRBP
 & \multirow[c]{3}{*}{$\bar d_v$}
 & \multirow[c]{3}{*}{$\bar d_v$}
 & \multirow[c]{3}{*}{$\bar d_v(\bar d_c-1)$}
 & \multirow[c]{3}{*}{$N-1$}
 & \multirow[c]{3}{*}{$\bar d_v(\bar d_c-1)$}
\\ \cline{1-1}
NW-E-CIRBP & & & & &
\\ \cline{1-1}
NW-Q-CIRBP & & & & &
\\ \hline
\end{tabular}
}
\end{table}

For the CIRBP, E-CIRBP, and Q-CIRBP decoders,
$N-1$ comparisons are first required to identify
the VN with the largest CI-related metric,
followed by one comparison with the corresponding threshold.
When the largest metric value is greater than or equal to the threshold,
the residual search is restricted to the incoming edges
of the selected VN,
resulting in $N+(\bar d_v-1)$ comparisons.
If the metric value is below the threshold,
a global residual search is performed instead
and additional comparisons are required.
The comparison requirement is therefore indicated by
$\geq N+(\bar d_v-1)$ in
Table~\ref{tab:per_update_complexity}.
For the arithmetic-operation accounting of the CI-related metrics,
additions and subtractions are assumed to have the same computational cost
and are collectively referred to as additions (ADDs).
Following the original CIRBP implementation \cite{CIRBP},
the posterior probabilities corresponding to the current
and precomputed total LLRs are obtained by table look-up.
Therefore, the exponential and division operations
in the direct evaluation of
(\ref{eq:posterior_probability})
are not included in the operation counts.
At each CI-related metric update,
only the quantity associated with the posterior probability
whose corresponding total LLR has changed
needs to be recomputed.
For the original CI metric,
one ADD is required to compute the difference
between the current and precomputed posterior probabilities.
For the E-CI metric,
computing the entropy complement
for the changed posterior probability requires,
in addition to the original CI computation,
three ADDs,
two multiplications (MULs),
and two logarithmic operations (LOGs).
The three additional ADDs are required to compute
$1-p$ for the changed posterior probability,
to combine the two terms in $H(p)$,
and to obtain the entropy complement $1-H(p)$.
For the Q-CI metric,
one additional ADD is required to compute $(p-0.5)$,
and one extra MUL is needed for the squared term.
The signs of $(p_n-0.5)$ and $(\widetilde p_n-0.5)$
in the evaluation of the E-CI and Q-CI metrics
can be directly obtained from the signs
of the corresponding total LLRs.
Hence, sign extraction, sign inversion,
and the final absolute-value operation
are assumed to be implemented through sign-bit manipulation
and are not included in the arithmetic-operation count.

The threshold setting of the edge-wise CI-based decoders
also results in a tradeoff between
the edge-selection complexity and FER performance.
As indicated in Table~\ref{tab:per_update_complexity},
when $\gamma_{\mathrm{E\text{-}CI}}=0$
or $\gamma_{\mathrm{Q\text{-}CI}}=0$,
the global maximum-residual edge search is avoided,
and the number of comparisons becomes
$N+(\bar d_v-1)$ per scheduling update.
Figures~\ref{ITEvsFER_MK1008_EWIDS_2.75dB} and
\ref{ITEvsFER_WIFI1944_EWIDS_1.75dB}
show that,
under these zero-threshold settings,
the E-CIRBP and Q-CIRBP decoders still outperform
the CIRBP decoder with its
optimized nonzero threshold
throughout the observed decoding iterations.
Therefore, for the M-1008 and W-1944 codes,
the global maximum-residual edge search can be avoided
while the proposed decoders still
provide better FER performance
than the CIRBP decoder using its optimized nonzero threshold.
For the N-3200 code, however,
the selected nonzero thresholds
provide better FER performance
than the corresponding zero-threshold settings,
as discussed above.
The threshold setting therefore affects both
the FER performance and the edge-selection complexity,
and the preferred setting can depend on the code being decoded.

\begin{table}[!htbp]
\centering
\caption{Average decoding iterations and normalized CI-related metric operation counts for the M-1008 code; $I_{\max}=35$.}
\label{tab:metric_m1008}
\begingroup
\small
\setlength{\tabcolsep}{5pt}
\renewcommand{\arraystretch}{1.1}
\begin{tabular}{|c|c|c|c|c|c|}
\hline
$E_b/N_0$ (dB) & Decoder & Average Iterations ($\bar I$) & ADDs & MULs & LOGs \\
\hline
\multirow{3}{*}{1.75} & NW-CIRBP & 3.7014 & 3.7014 & 0 & 0 \\
\cline{2-6}
 & NW-E-CIRBP & 2.8083 & 11.2332 & 5.6166 & 5.6166 \\
\cline{2-6}
 & NW-Q-CIRBP & 3.0781 & 6.1562 & 3.0781 & 0 \\
\hline
\multirow{3}{*}{2} & NW-CIRBP & 1.7845 & 1.7845 & 0 & 0 \\
\cline{2-6}
 & NW-E-CIRBP & 1.4378 & 5.7512 & 2.8756 & 2.8756 \\
\cline{2-6}
 & NW-Q-CIRBP & 1.4869 & 2.9738 & 1.4869 & 0 \\
\hline
\multirow{3}{*}{2.25} & NW-CIRBP & 1.1840 & 1.1840 & 0 & 0 \\
\cline{2-6}
 & NW-E-CIRBP & 1.0789 & 4.3156 & 2.1578 & 2.1578 \\
\cline{2-6}
 & NW-Q-CIRBP & 1.0925 & 2.1850 & 1.0925 & 0 \\
\hline
\end{tabular}
\endgroup
\end{table}

\begin{table}[!htbp]
\centering
\caption{Average decoding iterations and normalized CI-related metric operation counts for the W-1944 code; $I_{\max}=35$.}
\label{tab:metric_w1944}
\begingroup
\small
\setlength{\tabcolsep}{5pt}
\renewcommand{\arraystretch}{1.1}
\begin{tabular}{|c|c|c|c|c|c|}
\hline
$E_b/N_0$ (dB) & Decoder & Average Iterations ($\bar I$) & ADDs & MULs & LOGs \\
\hline
\multirow{3}{*}{1.375} & NW-CIRBP & 2.8756 & 2.8756 & 0 & 0 \\
\cline{2-6}
 & NW-E-CIRBP & 2.3979 & 9.5916 & 4.7958 & 4.7958 \\
\cline{2-6}
 & NW-Q-CIRBP & 2.4643 & 4.9286 & 2.4643 & 0 \\
\hline
\multirow{3}{*}{1.5} & NW-CIRBP & 2.3536 & 2.3536 & 0 & 0 \\
\cline{2-6}
 & NW-E-CIRBP & 2.1176 & 8.4704 & 4.2352 & 4.2352 \\
\cline{2-6}
 & NW-Q-CIRBP & 2.1500 & 4.3000 & 2.1500 & 0 \\
\hline
\multirow{3}{*}{1.625} & NW-CIRBP & 2.1060 & 2.1060 & 0 & 0 \\
\cline{2-6}
 & NW-E-CIRBP & 1.9648 & 7.8592 & 3.9296 & 3.9296 \\
\cline{2-6}
 & NW-Q-CIRBP & 1.9855 & 3.9710 & 1.9855 & 0 \\
\hline
\end{tabular}
\endgroup
\end{table}

To further compare the computational requirements
of the CI, E-CI, and Q-CI metrics
under the same node-wise decoding framework,
Tables~\ref{tab:metric_m1008} and \ref{tab:metric_w1944}
report the average decoding iterations (denoted by $\bar I$)
and the corresponding normalized CI-related metric operation counts
for the NW-CIRBP, NW-E-CIRBP, and NW-Q-CIRBP decoders
with $I_{\max}=35$.
For each frame,
decoding is terminated once
all parity-check equations are satisfied;
otherwise, it continues until $I_{\max}$ is reached.
To focus on the relative arithmetic costs of the three
CI-related metrics, the operation counts in
Tables~\ref{tab:metric_m1008} and~\ref{tab:metric_w1944}
are normalized on a per-metric-update basis for one VN.
As a result, for an average number
of decoding iterations $\bar I$,
the normalized metric-operation counts
are obtained by multiplying
$\bar I$ by the number of arithmetic operations required
for one CI-related metric update.
Specifically, the normalized metric evaluation
requires $\bar I$ ADDs for the NW-CIRBP decoder,
$4\bar I$ ADDs, $2\bar I$ MULs, and $2\bar I$ LOGs for the NW-E-CIRBP decoder,
and $2\bar I$ ADDs and $\bar I$ MULs for the NW-Q-CIRBP decoder.

As shown in Tables~\ref{tab:metric_m1008} and
\ref{tab:metric_w1944},
both the NW-E-CIRBP and NW-Q-CIRBP decoders
require fewer average decoding iterations
than the NW-CIRBP decoder
at all the tested $E_b/N_0$ values.
This is consistent with
the faster FER reduction
of the proposed metrics during the early decoding iterations
observed in
Figures~\ref{ITEvsFER_MK1008_NWIDS_2.75dB} and
\ref{ITEvsFER_WIFI1944_NWIDS_1.75dB}.
Since the three decoders use
the same node-wise updating procedure,
the reduced iteration counts can also reduce
the common decoding operations,
including C2V message propagations,
V2C updates,
C2V precomputations,
and related comparisons.
Among the three CI-based node-wise algorithms,
the NW-E-CIRBP decoder requires
the fewest average decoding iterations
at every tested $E_b/N_0$,
while the values obtained with the NW-Q-CIRBP decoder
remain close to those of NW-E-CIRBP.
The normalized ADD and MUL counts of NW-Q-CIRBP
are approximately one half of those of NW-E-CIRBP,
and NW-Q-CIRBP requires no LOG operation.
Therefore,
the Q-CI metric requires fewer metric-evaluation operations
than the E-CI metric,
while retaining a similar reduction
in the average number of decoding iterations.
Compared with the original CI metric, 
both proposed metrics have higher normalized CI-related 
arithmetic-operation counts, 
even after the reduction in average decoding iterations 
is taken into account.
On the other hand,
the NW-E-CIRBP and NW-Q-CIRBP decoders
provide better FER performance
and require fewer average decoding iterations
than the NW-CIRBP decoder
for the cases considered above.
The additional metric operations
are therefore traded for improved FER performance
and faster decoding convergence,
with the Q-CI metric providing
a lower-complexity alternative to the E-CI metric
while preserving most of its convergence improvement.

\section{Conclusions}
\label{Sec5}

In this paper,
we have proposed the E-CI and Q-CI metrics
to improve the original CI metric
used for determining the update priorities of VNs
in IDS-based BP decoding of LDPC codes.
The E-CI metric uses the binary entropy complement
to represent the certainty
of the current and precomputed posterior probabilities,
whereas the Q-CI metric is derived from
the leading quadratic term
in the local expansion of the entropy complement
to reduce the computational cost
associated with direct E-CI evaluation.
Both proposed metrics were incorporated into
the generalized CIRBP decoding procedure
and its node-wise extension,
allowing different metrics to be used
for determining the update priorities of VNs.
Simulation results for the M-1008, W-1944, and N-3200 codes
indicate that the proposed E-CI and Q-CI metrics
can generally provide lower FER
than the original CI metric
under the considered decoding procedures.
The relative FER advantages provided
by the E-CI and Q-CI metrics
depend on the code, the updating strategy,
and the number of decoding iterations.
In addition to their FER performance,
the two proposed metrics also differ
in the computational cost of
evaluating the metrics themselves,
with the Q-CI metric avoiding
the logarithmic operations required
for direct E-CI evaluation.
Although both metrics require more computational effort
for each metric evaluation than the original CI metric,
the improved early-iteration decoding performance obtained
with the proposed metrics indicates the potential for faster
decoding convergence.
This behavior is further supported by the complexity analysis
for the M-1008 and W-1944 codes, where the node-wise decoders
employing the proposed metrics require fewer average decoding
iterations than the NW-CIRBP decoder.
Therefore, when the costs of metric evaluation and the decoding
convergence behavior are considered jointly, the additional
computational cost of metric evaluation may be partly balanced by
the reduction in the average number of decoding iterations.
The results reported in this paper further suggest that
considering the individual reliability levels
associated with the current and precomputed posterior probabilities,
rather than relying solely on
their absolute difference, may provide additional discriminative
information for designing metrics that prioritize VNs for updating.
Future work may investigate whether other characteristics of the
posterior probabilities can be utilized to further enhance
VN-selection metric design.

Two practical directions also deserve further investigation.
One is how different scheduling metrics and updating strategies
for IDS-based BP decoding can be adapted
to practical block-structured LDPC codes,
such as QC-LDPC codes used in Wi-Fi and 5G NR,
including the development of suitable block-level metrics and
block-parallel scheduling rules.
In such an extension, the VN-level metric values
within a block
would need to be appropriately
combined to determine block-level priorities,
while the message-updating strategy within and across the selected blocks
would also need to be designed.
Different choices may result in different tradeoffs among
decoding performance, computational complexity, and parallelism.

The other practical direction concerns finite-precision implementation
of IDS-based decoding and the associated scheduling metrics.
For the proposed E-CI and Q-CI metrics,
quantization can be introduced at different stages
of the decoding and metric computations,
including the LLR-related quantities,
the current and precomputed posterior probabilities,
the quantities involved in metric evaluation,
and the resulting metric values.
Since these quantities have different numerical ranges
and are involved in different numerical operations,
the appropriate quantization design can differ across these stages.
Such implementation choices may affect the computed metric values,
the resulting candidate ordering,
and ultimately the decoding performance.
Future work could therefore investigate
finite-precision implementations of different scheduling metrics
for IDS-based BP decoding
under systematically designed quantization settings.


\begin{thebibliography}{999}

\bibitem[Gallager(1963)]{LDPC}
Gallager, R.G. \textit{Low-Density Parity-Check Codes}; MIT Press: Cambridge, MA, USA, 1963.

\bibitem[IEEE(2016)]{WiFi}
IEEE. \textit{IEEE Standard for Information Technology--Telecommunications and Information Exchange Between Systems Local and Metropolitan Area Networks--Specific Requirements--Part 11: Wireless LAN Medium Access Control (MAC) and Physical Layer (PHY) Specifications}; IEEE Std 802.11-2016; IEEE: New York, NY, USA, 2016.

\bibitem[3GPP(2018)]{NR}
3GPP. \textit{5G; NR; Multiplexing and Channel Coding}; Technical Specification (TS) 38.212, Release 15; 3rd Generation Partnership Project (3GPP): Sophia Antipolis, France, 2018.

\bibitem[ETSI(2014)]{DVBS2}
ETSI. \textit{Digital Video Broadcasting (DVB); Second Generation Framing Structure, Channel Coding and Modulation Systems for Broadcasting, Interactive Services, News Gathering and Other Broadband Satellite Applications; Part 1: DVB-S2}; ETSI EN 302 307-1 V1.4.1; ETSI: Sophia Antipolis, France, 2014.

\bibitem[Hocevar(2004)]{LBP}
Hocevar, D.E. A reduced complexity decoder architecture via layered decoding of LDPC codes. In Proceedings of the 2004 IEEE Workshop on Signal Processing Systems (SIPS 2004), Austin, TX, USA, 13--15 October 2004; pp. 107--112.

\bibitem[Zhang and Fossorier(2005)]{SBP}
Zhang, J.; Fossorier, M.P.C. Shuffled iterative decoding. {\em IEEE Trans. Commun.} {\bf 2005}, {\em 53}, 209--213.

\bibitem[Chang et al.(2009)]{newGHSBP}
Chang, C.-Y.; Chen, Y.-L.; Lee, C.-M.; Su, Y.T. New group shuffled BP decoding algorithms for LDPC codes. In Proceedings of the 2009 IEEE International Symposium on Information Theory (ISIT), Seoul, Republic of Korea, 28 June--3 July 2009; pp. 1664--1668.

\bibitem[Kim et al.(2012)]{Row_Layer}
Kim, M.; Kim, D.; Lee, Y.H. Serial scheduling algorithm of LDPC decoding for multimedia transmission. In Proceedings of the 2012 IEEE International Symposium on Broadband Multimedia Systems and Broadcasting (BMSB), Seoul, Republic of Korea, 27--29 June 2012; pp. 1--4.

\bibitem[Wang et al.(2020)]{LPHD_RB-LBP}
Wang, B.; Zhu, Y.; Kang, J. Two effective scheduling schemes for layered belief propagation of 5G LDPC codes. {\em IEEE Commun. Lett.} {\bf 2020}, {\em 24}, 1683--1686.

\bibitem[Manada et al.(2016)]{LG-GSBP}
Manada, A.; Yoshida, K.; Morita, H.; Tatsukawa, R. A grouping based on local girths for the group shuffled belief propagation decoding. {\em IEEE Commun. Lett.} {\bf 2016}, {\em 20}, 2133--2136.

\bibitem[Chang and Su(2017)]{AGSBP}
Chang, T.C.-Y.; Su, Y.T. Adaptive group shuffled decoding for LDPC codes. {\em IEEE Commun. Lett.} {\bf 2017}, {\em 21}, 2118--2121.

\bibitem[Touati and de Lamare(2024)]{WRB-LBP}
Touati, H.; de Lamare, R.C. Weighted residual layered belief propagation for LDPC decoding. In Proceedings of the 2024 19th International Symposium on Wireless Communication Systems (ISWCS), Rio de Janeiro, Brazil, 14--17 July 2024; pp. 1--6.

\bibitem[Vila Casado et al.(2007)]{RBP_NWRBP}
Vila Casado, A.I.; Griot, M.; Wesel, R.D. Informed dynamic scheduling for belief-propagation decoding of LDPC codes. In Proceedings of the 2007 IEEE International Conference on Communications (ICC), Glasgow, UK, 24--28 June 2007; pp. 932--937.

\bibitem[Lee et al.(2013)]{SVNF}
Lee, H.-C.; Ueng, Y.-L.; Yeh, S.-M.; Weng, W.-Y. Two informed dynamic scheduling strategies for iterative LDPC decoders. {\em IEEE Trans. Commun.} {\bf 2013}, {\em 61}, 886--896.

\bibitem[Aslam et al.(2017)]{DSVNFS}
Aslam, C.A.; Guan, Y.L.; Cai, K.; Han, G. Low-complexity belief-propagation decoding via dynamic silent-variable-node-free scheduling. {\em IEEE Commun. Lett.} {\bf 2017}, {\em 21}, 28--31.

\bibitem[Zhang and Chen(2019)]{RDRBP}
Zhang, H.; Chen, S. Residual-decaying-based informed dynamic scheduling for belief-propagation decoding of LDPC codes. {\em IEEE Access} {\bf 2019}, {\em 7}, 23656--23666.

\bibitem[Chang et al.(2022)]{UBPRBP}
Chang, T.C.-Y.; Lee, I.-H.; Wang, P.-H.; Weng, J.-J.; Su, Y.T. Enhanced informed dynamic BP decoding scheduling strategies for 5G NR LDPC codes. In Proceedings of the 2022 IEEE 96th Vehicular Technology Conference (VTC2022-Fall), London, UK, 26--29 September 2022; pp. 1--6.

\bibitem[Gong et al.(2011)]{IVCRBP}
Gong, Y.; Liu, X.; Ye, W.; Han, G. Effective informed dynamic scheduling for belief propagation decoding of LDPC codes. {\em IEEE Trans. Commun.} {\bf 2011}, {\em 59}, 2683--2691.

\bibitem[Liu et al.(2015)]{OVRBP}
Liu, X.; Zhang, Y.; Cui, R. Variable-node-based dynamic scheduling strategy for belief-propagation decoding of LDPC codes. {\em IEEE Commun. Lett.} {\bf 2015}, {\em 19}, 147--150.

\bibitem[Liu et al.(2016)]{V-VCRBP}
Liu, X.; Zhou, Z.; Cui, R.; Liu, E. Informed decoding algorithms of LDPC codes based on dynamic selection strategy. {\em IEEE Trans. Commun.} {\bf 2016}, {\em 64}, 1357--1366.

\bibitem[Liu et al.(2019)]{TRM_RM}
Liu, X.; Zi, L.; Yang, D.; Wang, Z. Improved decoding algorithms of LDPC codes based on reliability metrics of variable nodes. {\em IEEE Access} {\bf 2019}, {\em 7}, 35769--35778.

\bibitem[Chang et al.(2021)]{CIRBP}
Chang, T.C.-Y.; Wang, P.-H.; Weng, J.-J.; Lee, I.-H.; Su, Y.T. Belief-propagation decoding of LDPC codes with variable node--centric dynamic schedules. {\em IEEE Trans. Commun.} {\bf 2021}, {\em 69}, 5014--5027.

\bibitem[MacKay(n.d.)]{MacKayWeb}
MacKay, D.J.C. \textit{Encyclopedia of Sparse Graph Codes}. Available online: \url{http://www.inference.phy.cam.ac.uk/mackay/codes/data.html} (accessed on 3 September 2026).

\end{thebibliography}
\end{document}